\documentclass[twocolumn,aps]{revtex4-1}
\usepackage{graphicx}
\usepackage{amssymb}
\usepackage{amsfonts}
\usepackage{amsmath}
\usepackage{amsbsy}
\usepackage{natbib}
\usepackage{comment}

\usepackage{color}
\usepackage{float}

\usepackage[colorlinks=true, urlcolor=blue, citecolor=blue, linkcolor=blue]{hyperref}

\renewcommand{\vec}[1]{\mbox{\boldmath$\mathrm{#1}$}}
\let\sb=_ \catcode`\_=\active \def_#1{\ensuremath \sb{\rm#1}}

\renewcommand{\vec}[1]{\mbox{\boldmath$\mathrm{#1}$}}
\newcommand{\be}{\begin{equation}}
\newcommand{\ee}{\end{equation}}
\newcommand{\ben}{\begin{eqnarray}}
\newcommand{\een}{\end{eqnarray}}

\begin{document}

%\preprint{APS/123-QED}

\title{Thermally assisted Skyrmion drag in a nonuniform electric field}

\author{Xi-guang Wang$^{1,2}$, L. Chotorlishvili$^{2}$,  Guang-hua Guo$^1$, C.-L. Jia$^3$, J. Berakdar$^2$}

\address{$^1$School of Physics and Electronics, Central South University, Changsha 410083, China\\
$^2$Institut f\"ur Physik, Martin-Luther Universit\"at Halle-Wittenberg, 06099 Halle/Saale, Germany\\
$^3$ Key Laboratory for Magnetism and Magnetic Materials of the Ministry of Education, Lanzhou University, Lanzhou 730000, China}

\date{\today}% It is always \today, today,

\begin{abstract}
Magnetic skyrmions are topologically protected excitations of the magnetization vector field with promising applications in spintronics and spin-caloritronics, particularly due to their high mobility. Skyrmions can be steered by  a spin-polarized charge current or by exposure to a magnonic spin current.
Here, we propose a further method for driving skyrmions by applying an inhomogeneous electric field and a homogeneous thermal bias.
We show that the inhomogeneous electric torque leads to an efficient skyrmionic drag which can be thermally assisted as to enhance the skyrmion velocity.
The calculations and analysis  are limited to  insulating samples; for conducting materials the influence of the inhomogeneous electric field on the charge carriers need to be taken also into account.

\end{abstract}

\maketitle

\section{Introduction}

The notion of  skyrmion  was introduced in high-energy physics by Skyrme \cite{Skyrme} as a topologically stable vector field configuration. The vector order parameter of a magnetically ordered system also  shows  such stable excitations, as experimentally
observed in  the chiral itinerant compound MnSi  \cite{Binz}.
Essential for the formation of magnetic skyrmions  is the absence of inversion symmetry \cite{Iwasaki,Dai,HoonHan,Bogdanov,Garst,Gavilano}, that can be imposed
for instance by the Dzyaloshinskii-Moriya interaction.
Skyrmions exhibit a quasi-particle character with mobility higher than a magnetic domain wall \cite{Bulaevskii, Kong, Mishra,Batista,Hoogdalem, Papanicolaou}. Besides,
  they are less pinned to dislocations and impurities and may have an extension on the nanometer scale,  which make them interesting for spintronic applications
\cite{Saxena, Rosch,Finocchio}. An important issue thereby is the controlled driving of skyrmions. One may set skyrmion in motion using a spin-polarized charge current \cite{Jonietz,Iwasaki1463,yzhou2016} or as a thermally assisted magnonic current \cite{Batista, Kong, Mochizuki2014}.

The pressure exerted by spin transfer torque (STT) on the surface of skyrmion moves the skyrmion and allows achieving reasonably high speed.
The magnonic spin Seebeck current exploits the presence of a thermal gradient which is usually difficult to realize and hard to control swiftly on the nanoscale.
To generate STT, various types of methods are employed, including the spin Hall effect in heavy metal/magnet heterostructure \cite{Liu2012,sinova2015}, spin injection in nonlocal structure \cite{Demidov2016, Lin5177}, and STT in metallic multilayer \cite{Slonczewski, Berger}. Each of these methods provides a unique insight into the spin transfer torque manipulation. Nevertheless, standard methods require demanding preparation of the sample, and their application is usually hampered by Joule heating or interface effect.

In the present work, we show that at a finite uniform temperature, applied inhomogeneous electric field generates an inhomogeneous electric torque (IET) that can be used to drive skyrmions. Such fields are widely available, and their spatiotemporal structure is accurately controllable. The proposed scheme 
	does not rely on dissipative charge currents and hence involves less energy dissipation. In addition, electric fields are more easily spatio-temporally  engineered and controlled on the nanoscale than  temperature gradients that might be also used to move skyrmions. One should note however,
	 that in principle, an inhomogeneous electric field applied to an electrical conductor induces a local inhomogeneous Joule heating and through the formed inhomogeneous temperature profile can influence the magnon density.  Thus, the inhomogeneous Joule heating has an extra effect that can additionally influence  the skyrmion drag but further dissipates energy.   This effect is absent for insulating materials to which the current study is restricted.

The paper is organized as follows: in Sec. \ref{sec:model}  we specify the model, in Sec. \ref{IET}  we explore the mechanisms of the formation of IET, in Sec. \ref{damping} we study the inhomogeneous damping and the intrinsic frequencies of the system. The thermally assisted magnonic current is addressed in the Sec. \ref{sec:spincurrent}, in the Sec. \ref{sec:motion}  we analyze the skyrmion motion induced by the IET. The final Sec. \ref{pulse}
is dedicated to the effect of the time-dependent electric field.

\section{Model}
\label{sec:model}
In spite of the absence of itinerant electrons in single phase multiferroic  or  magnetic insulators,
the virtual hopping of electrons between the $d$ orbitals and the strong spin-orbit interaction leads to a
net ferroelectric polarization $\vec{P} =  c_{E} [(\vec{m} \cdot \nabla) \vec{m} - \vec{m} (\nabla \cdot \vec{m}) ]$. Here, $ \vec{m} $
is the unit magnetization vector of the magnet, and $ c_E $ is a magneto-electric (ME) coupling constant (see \cite{Tliu20,VRis} for details).
The net ferroelectric polarization is coupled to the applied external electric field $ \vec{E} = (E_x, E_y, E_z) $. The ME coupling energy $ E_{ele} = - \vec{E} \cdot \vec{P} $
mimics that resulting from  a dynamical Dzyaloshinskii-Moriya (DM) interaction  and leads to the formation of N$ \mathrm{\acute{e}} $el-type skyrmion .
The $E_z$ component of the field,
applied along the whole sample stabilizes the skyrmion \cite{Troncoso2016, loidl2015, Tokura2013}. The same skyrmion structure can be stabilized utilizing the bulk (in the material with broken inversion symmetry in lattices) or interface type DM interaction term (at the interface of magnetic films) \cite{Tokura2013}.

For a  uniform finite  temperature  the magnetization dynamics is governed by the stochastic Landau-Lifshitz-Gilbert (LLG) equation supplemented by the ME term
\begin{equation}
\displaystyle \frac{\partial \vec{M}}{\partial t} = - \gamma \vec{M} \times \bigg(\vec{H}_{\mathrm{eff}} +\vec{h}_{l} - \frac{1}{\mu_0 M_s } \frac{\delta E_{ele}}{\delta \vec{m}}\bigg)+ \frac{\alpha}{M_{s}} \vec{M} \times \frac{\partial \vec{M}}{\partial t}.
\label{LLG}
\end{equation}
Here, $\vec{M} =  M_{s}\vec{m} $ and $M_{s}$ is the saturation magnetization,  $ \gamma $ is the gyromagnetic ratio, and $ \alpha $ is the phenomenological Gilbert damping constant. The effective field $ \vec{H}_{\mathrm{eff}} $ consists of the exchange field and the applied external magnetic filed, $ \vec{H}_{\mathrm{eff}} = \frac{2 A_{ex}}{\mu_0 M_s} \nabla^2 \vec{m} + H_z \vec{z} $, where $ A_{ex} $ is the exchange stiffness, and $ H_z $ is the external magnetic field applied along the \textit{z}-direction. The thermal random magnetic field is characterized by the correlation function of a white noise \cite{JLGar22}, meaning  $   \langle \vec{h}_{l,i}(x,t) \vec{h}_{l,j}(x',t') \rangle = \frac{2 k_{B} T \alpha}{\gamma M_{s} V}\delta_{ij} \delta(x-x') \delta(t-t') $ , where $ k_B $ is the Boltzmann constant and $ T $ is the temperature.

Existence of skyrmions can be verified through the topological number $ Q = \int dxdy \rho_{\mathrm{sky}} $, with $ \rho_{\mathrm{sky}} = -\frac{1}{4\pi} \vec{m} \cdot (\partial_x \vec{m} \times \partial_y \vec{m}) $. In the case of  a single skyrmion, we obtain $ Q = 1 $. The inhomogeneous electric (E) field $ \left| E_y \right| $ is  applied only to a   part of the system. This can be achieved for instance a screening metalic layer to block the E-field from the respective region. In what follows we show that the inhomogeneous E-field $ E_y $  modifies the thermal magnon density profile and the induced magnon flow drives the skyrmion. The numerical simulations based on Eq. (\ref{LLG}) are done at zero and finite temperatures for the saturation magnetization $ M_s = 1.4 \times 10^5$ A/m, the exchange constant $ A_{ex} = 3 \times 10^{-12} $ J/m, the ME coupling strength $ c_E = 0.9 $ pC/m, and the Gilbert damping constant $ \alpha = 0.001$. The N$ \mathrm{\acute{e}} $el-type skyrmion is stabilized by the electric and magnetic fields $ E_z = 1.7 $ MV/cm,  $ H_z = 3.2 \times 10^5 $ A/m. The transversal component of the electric field $ \left| E_y \right|  $ is in the order of (0, 0.15 MV/cm):  This value of $ \left| E_y \right|  $ is small enough and cannot induce switching of the equilibrium magnetization.

\section{Inhomogeneous electric torque}
\label{IET}

The term $\frac{1}{\mu_0 M_s } \frac{\delta E_{ele}}{\delta \vec{m}}$ that enters in the LLG equation Eq.(\ref{LLG}) quantifies the influence  of the ME coupling on the  effective  field. Thus,  the ME coupling constant $c_E$ and the configuration of the applied electric field are important issues to consider. We investigated  inhomogeneous electric field $ \vec{E}(x) $ (varying along the $ x $ axis) leading to
\begin{equation}
\begin{small}
\begin{aligned}
-\frac{1}{\mu_0 M_s}\frac{\delta E_{ele}(E_i)}{ \delta \vec{m}} = &\frac{c_E}{\mu_0 M_s} [\partial_x E_i (m_i \vec{e}_x-m_x\vec{e}_i) \\
                                 & + \sum_{j(j \ne i)}2E_i(-\partial_j m_j \vec{e}_i + \partial_j m_i \vec{e}_j)].
\label{IET1}
\end{aligned}
\end{small}
\end{equation}
Here, $ i,j = x, y, z$. The total ME torque term that enters  the LLG equation $  -\gamma \vec{m} \times [-\frac{1}{\mu_0 M_s}\frac{\delta E_{ele}}{ \delta \vec{m}}]  $ has two sources: the electric field $ \vec{E} = (E_x, E_y, E_z) $ and its gradient $ \partial_x \vec{E} = (\partial_x E_x, \partial_x E_y, \partial_x E_z)  $. The inhomogeneity of the electric field is manifested in the spatially-inhomogeneous DM interaction and in the additional  terms $\partial_x E_i $, where $ i = x,y ,z$.
After some algebra, we infer the expression for the IET solely induced by the electric field gradient $ \partial_x \vec{E} $
\begin{equation}
\displaystyle -\gamma \vec{m} \times \bigg(-\frac{\delta E_{ele}(\partial_x E_i)}{\mu_0 M_s \delta \vec{m}}\bigg) = -\frac{\gamma c_E \partial_xE_i}{\mu_0 M_s} \vec{m} \times (\vec{m} \times \vec{p_E}).
\label{IET2}
\end{equation}
The vector $ \vec{p_E} = \vec{x} \times \vec{e}_i$ is set by  $ \vec{e}_i $ which points into the direction of electric field.
Obviously  the expression Eq.(\ref{IET2}) is identical to the standard spin transfer torque $ -c_j \vec{m} \times (\vec{m} \times \vec{p}) $,  in which case  $\vec{p_E} $ mimics the spin polarization direction $\vec{p}$.
While $c_j$ depends on the electric current density, the amplitude of the IET depends on the gradient of an electric field $\partial_x E_i$ and the ME coupling strength $c_{E}$.

For further insight let us utilize an atomistic model in which the polarization vector
 $ \vec{P} $ is expressed as $ \vec{P} = -\frac{J e a}{E_{so}}\vec{e}_{n,n+1} \times (\vec{S}^n \times \vec{S}^{n+1}) $ , where
 $ \vec{S}^n $ is the spin  localized on the n-th site, the strength of the spin-orbit interaction is quantified by $ E_{SO} = \hbar^2 / (2m_e \lambda^2) $, $ m_e $ is the electron mass, $ \lambda $ is the spin-orbit coupling constant, $ J $ is the exchange coefficient, $ a $ is the distance between neighboring magnetic ions, $ e $ is the electron charge, $ \vec{e}_{n,n+1} $ is the unit vector connecting the ions.
The ME coupling term $ E_{ele}(\vec{S}^n) $ in the atomistic model can be rewritten in the form
\begin{equation}
\displaystyle E_{ele}(\vec{S}^n) = - \vec{E}^{n-1} \cdot \vec{P}^{n-1} - \vec{E}^{n} \cdot \vec{P}^{n} .
\label{IET3}
\end{equation}
We adopt the geometry of 1D chain stretched out along the $x$ axis  and for the electric field components $E_y,~E_z$ we write down the explicit expressions of ME terms:
\begin{equation}
\begin{small}
\begin{aligned}
\displaystyle E_{ele}(\vec{S}^n, E_y) &= -\frac{J e a}{E_{so}}[E_{y,0}(2S_x^n\Delta S_y - 2 S_y^n\Delta S_x) \\
                               & - \Delta E_y(S_x^n \bar{S}_y^{n} - S_y^n\bar{S}_x^{n})],\\
              E_{ele}(\vec{S}^n, E_z) &= -\frac{J e a}{E_{so}}[E_{z,0}(2 S_x^n\Delta S_z-2S_z^n\Delta S_x) \\
                              & - \Delta E_z(S_x^n\bar{S}_z^{n}-S_z^n \bar{S}_x^{n})].\\
\end{aligned}
\label{IET4}
\end{small}
\end{equation}
Here, we implemented the ansatz: $ \vec{E}_{n-1} = \vec{E}_{0} -\Delta \vec{E}/2$, $ \vec{E}_{n} = \vec{E}_{0} + \Delta\vec{E}/2$, and $ \vec{S}^{n-1} $ =  $ \langle\vec{S}^{n}\rangle - \Delta\vec{S}/2$, $ \vec{S}^{n+1} $ =  $ \langle\vec{S}^{n}\rangle + \Delta\vec{S}/2$, where $\langle\vec{S}^{n}\rangle$ means the averaged value of the spin. The atomistic IET has the form:
\begin{equation}
\displaystyle - \vec{S}^n \times \bigg(-\frac{\delta E_{ele}^{n}}{\delta \vec{S}^n}\bigg) = -\frac{e a J  \Delta E_i}{E_{so}} \vec{S}^n \times [\langle\vec{S}^n\rangle \times \vec{p_E} ],
\label{IET5}
\end{equation}
where, $ \vec{p_E} = \vec{e}_{n,n+1} \times \vec{e}_i$ and $\vec{e}_{n,n+1}$ is the unit vector connecting the sites $n,n+1$. In the continuous limit, the atomistic model Eq.(\ref{IET5})
goes over into  the continuum model upon  coarse graining Eq. (\ref{IET2}).
We use the following boundary conditions  for two-dimensional model \cite{Rohart2013}
\begin{equation}
\begin{aligned}
\displaystyle  \frac{\partial m_x}{\partial x} |_{\partial V} + \frac{c_E E_y}{2 A_{ex}}m_y+ \frac{c_E E_z}{2 A_{ex}}m_z &= 0,\\
\frac{\partial m_y}{\partial x} |_{\partial V} - \frac{c_E E_y}{2 A_{ex}}m_x &= 0,\\
\frac{\partial m_z}{\partial x} |_{\partial V} - \frac{c_E E_z}{2 A_{ex}}m_x &= 0,\\
\frac{\partial m_x}{\partial y} |_{\partial V} - \frac{c_E E_x}{2 A_{ex}}m_y &= 0,\\
\frac{\partial m_y}{\partial y} |_{\partial V} + \frac{c_E E_x}{2 A_{ex}}m_x + \frac{c_E E_z}{2 A_{ex}}m_z &= 0,\\
\frac{\partial m_z}{\partial y} |_{\partial V} - \frac{c_E E_z}{2 A_{ex}}m_y &= 0.\\
\end{aligned}
\label{boundary}
\end{equation}
Here, $ E_x $, $ E_y $ and $ E_z $ are boundary values of the electric field components. Derivatives in Eq. (\ref{LLG}) are implemented in the sense of central derivatives, i.e., $ \partial{\vec{m}}/\partial i = (\vec{m}_{ni+1} - \vec{m}_{ni-1}) / (2  \triangle_i ) $, where $ i = x, y $, and $ \triangle_i $ is the step size along the $ i $ direction.
Thus, beyond the total ME torque (including IET), the boundary conditions impose extra torque $ \vec{T}_{\mathrm{boundary}} $. However, while ME torque influences
the magnetization dynamics in the whole system, the extra torque imposed by boundary conditions is local and acts in the vicinity of boundaries.
In what follows the boundary torque we call the  "boundary magnetoelectric torque."

\section{damping}
\label{damping}

\begin{figure}
    \includegraphics[width=0.48\textwidth]{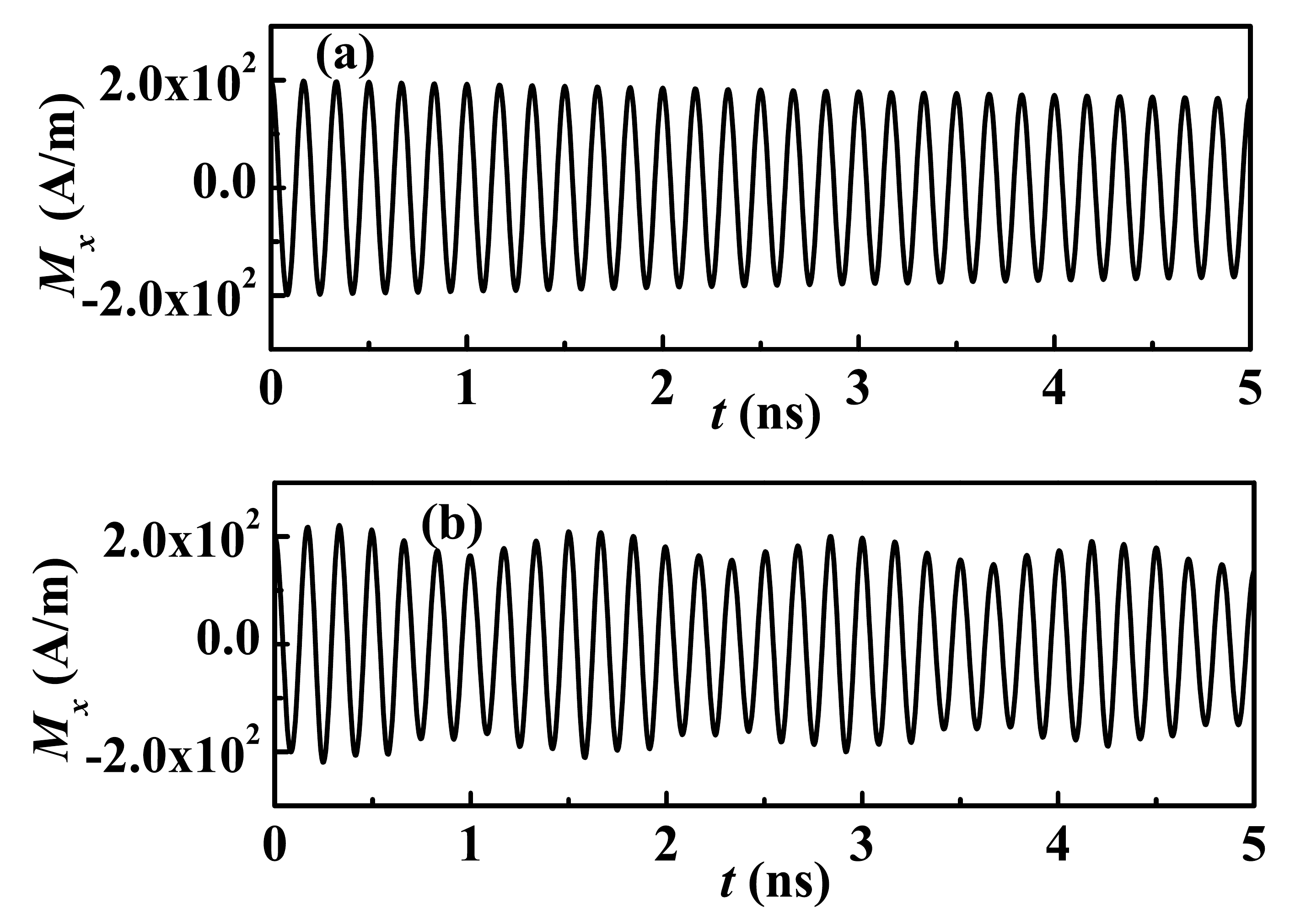}
    \caption{\label{Mx-time-E} The time dependence of the magnetization component $ M_x $ excited by a uniform perturbation.  Oscillations correspond to the electric field  $ \vec{E} = (0,0,0) $ (a), and $ \vec{E} = (0,E_{gr} x \vec{e}_y,0) $ (b). The E-field gradient is constant and is  chosen as $ E_{gr} = -0.24 $ (MV/m)/nm. In the simulations, the two-dimensional magnetic layer is located in the region of $ -125 \text{nm} \le x \le 125 $ nm and $ -125 \text{nm} \le y \le 125 $ nm.}
\end{figure}

\begin{figure}
    \includegraphics[width=0.48\textwidth]{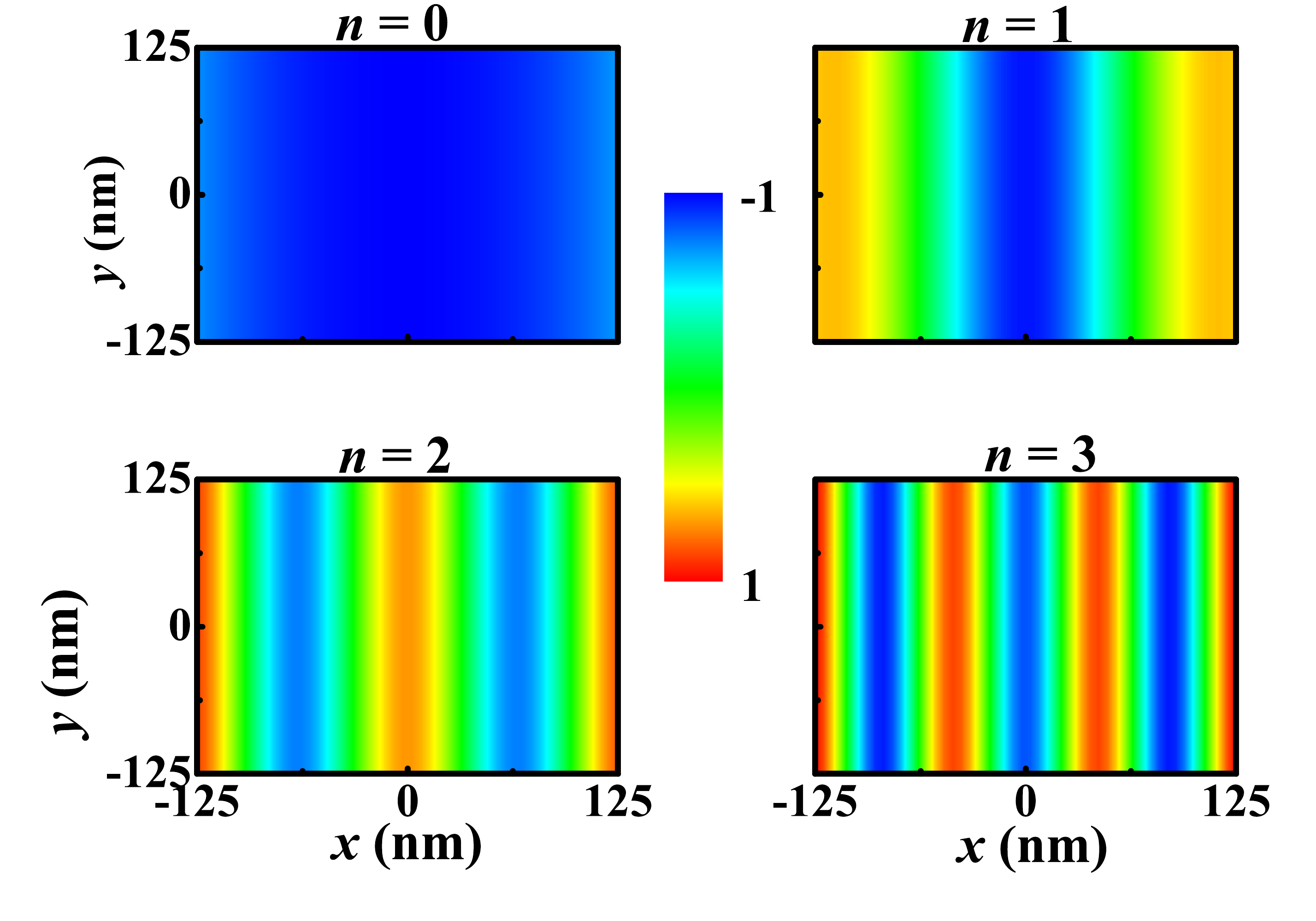}
    \caption{\label{peaks} Spatial distributions of four standing waves with $ 2n $ nodes ($ n=0,1,2 $ and 3) along the $ x $ axis.}
\end{figure}

From an energy point of view, the ME interaction provides an additional route for energy transfer. The ME interaction affects the magnetization precession as well as the damping constant $ \alpha_{\mathrm{eff}}$.  We inspect these effects based on the magnetic resonance analysis.
The time dependence of the magnetization $ \vec{M}(t) $ with or without the electric field is calculated based on the LLG equation (Eq. \ref{LLG}). For $ H_z = 1.6 \times 10^5 $ and a zero E field $ \vec{E} = (0,0,0) $, we estimate the resonance frequency and the relaxation time, respectively $ \omega_0 = \gamma H_0 = 2 \pi \times 6 $ GHz,  $ \tau_0 = 1/(\alpha \omega_0) = 26.5$ ns.
Note that, for the adopted excitations, only several intrinsic eigenmodes can be excited, for example, the ferromagnetic resonance and standing wave modes.

\begin{figure}
    \includegraphics[width=0.48\textwidth]{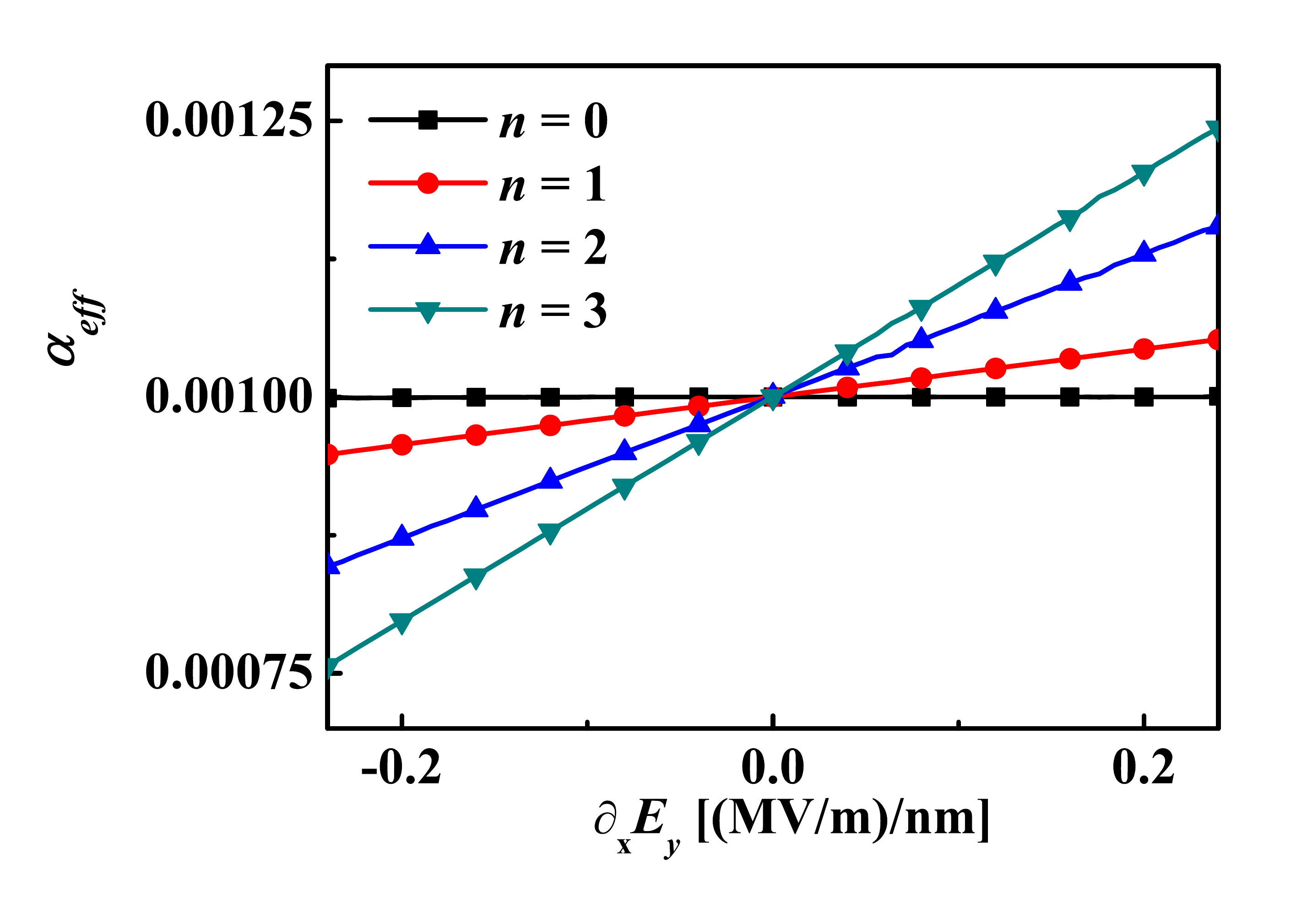}
    \caption{\label{alpha} The effective damping constant $ \alpha_{\rm {eff}} $ as a function of the inhomogeneous electric field $ \partial_xE_y $ for different standing waves with $ 2n $ modes ($ n=0,1,2 $ and 3).}
\end{figure}

As is shown in Fig. \ref{Mx-time-E}(a), only the single frequency $\omega = \omega_0 =6$ GHz magnetization oscillation is excited in the absence of the external electric field and the corresponding effective damping is constant $ \alpha_{\mathrm{eff}} = \alpha $. However, applying a linear electric field $ \vec{E} = E_{gr} x \vec{e}_y $ with $ E_{gr} = -0.24 $ (MV/m)/nm along the $ y $-direction (i.e., $ \vec{p}_E = \vec{e}_z $), changes the free-energy landscape  allowing the  activation of several high frequency modes, as  shown by Fig. \ref{Mx-time-E}(b). A detailed analysis  suggests that four frequencies $ \omega_1 = 2 \pi \times 5.99 $ GHz, $ \omega_2 = 2 \pi \times 6.75 $ GHz, $ \omega_3 = 2 \pi \times 9 $ GHz and $ \omega_4  = 2 \pi \times 12.73$ GHz dominate in Fig. \ref{Mx-time-E}(b). The spatial distributions of the dominant frequencies are presented in Fig. \ref{peaks}, and one can see that the dominant frequencies correspond to the standing waves with $ 2n $ nodes ($ n=0,1,2 $ and 3) along the $ x $ axis. Using the fitting equation $ M_x(t) = \sum_{i=1,2,3,4} A_i  \cos(\omega_it + \phi_i)\exp(-t/\tau_i)  $, we determine $ \tau_1 = 26.6 $ ns ($ \alpha_{\mathrm{eff}1} = 0.00099$), $ \tau_2 = 24.9 $ ns ($ \alpha_{\mathrm{eff}2} = 0.00095$), $ \tau_3 = 20.9 $ ns ($ \alpha_{\mathrm{eff}3} = 0.00085$) and $ \tau_4 = 16.5 $ ns ($ \alpha_{\mathrm{eff}4} = 0.00076$).
The applied uniform electric field induces these oscillation modes with $ 2n $ nodes and leaves the effective damping $ \alpha_{\mathrm{eff}} $ unchanged. However,
the effective damping $ \alpha_{\rm {eff}} $  depends linearly on the gradient of electric field $ \partial_xE_y $, see Fig. \ref{alpha}. Therefore, the effective damping can be controlled through an inhomogeneous
electric field.

\begin{figure}
    \includegraphics[width=0.48\textwidth]{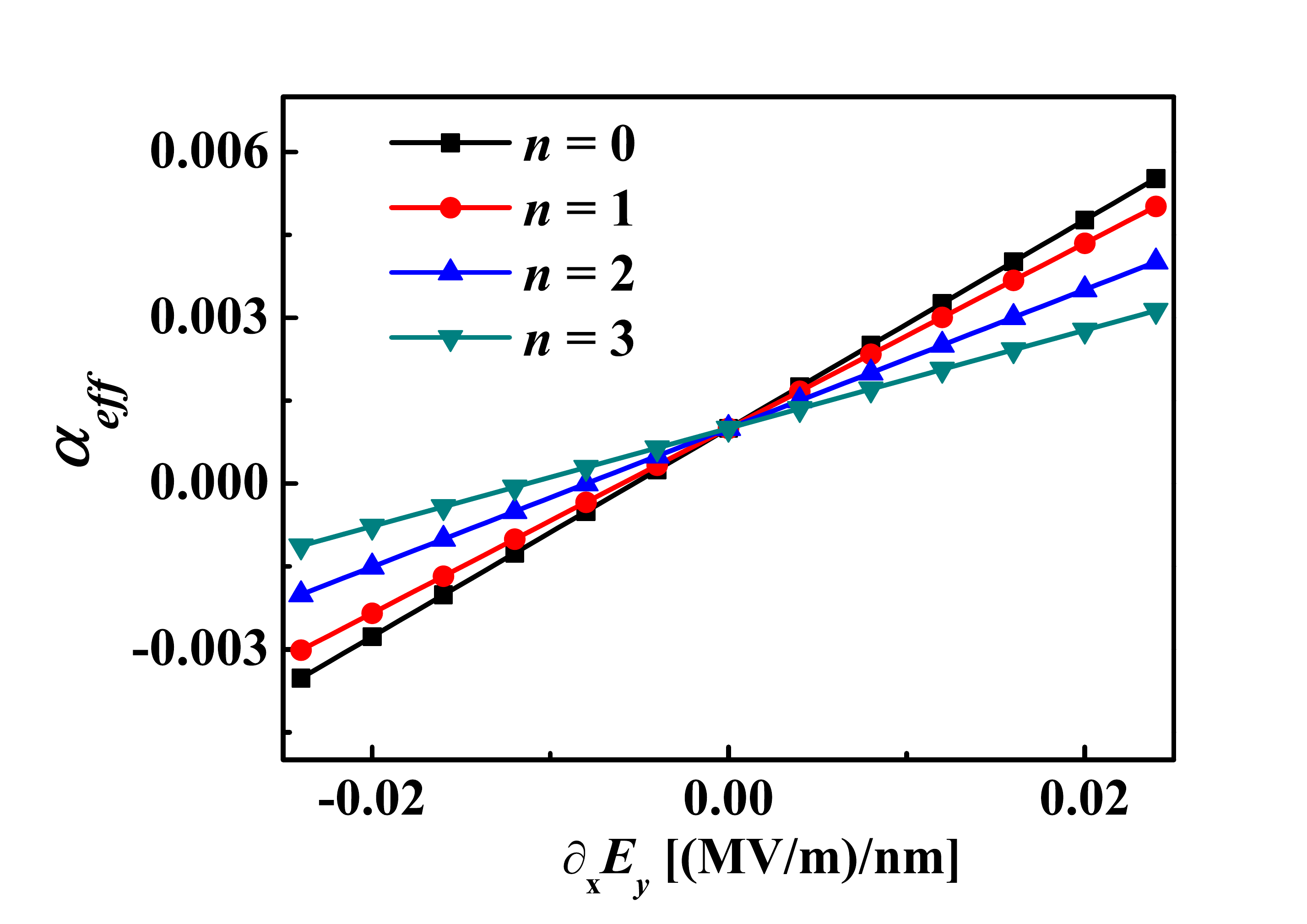}
    \caption{\label{alpha-STT} The roles of  boundary constraints and  boundary torque $ T_{\mathrm{boundary}} $ is illustrated by  excluding both while running the simulations.  The effective damping constant $ \alpha_{\rm {eff}} $ is plotted as a function of the electric field gradient for the different standing waves with $ 2n $ modes ($ n=0,1,2 $ and 3).}
\end{figure}

\begin{figure}
    \includegraphics[width=0.48\textwidth]{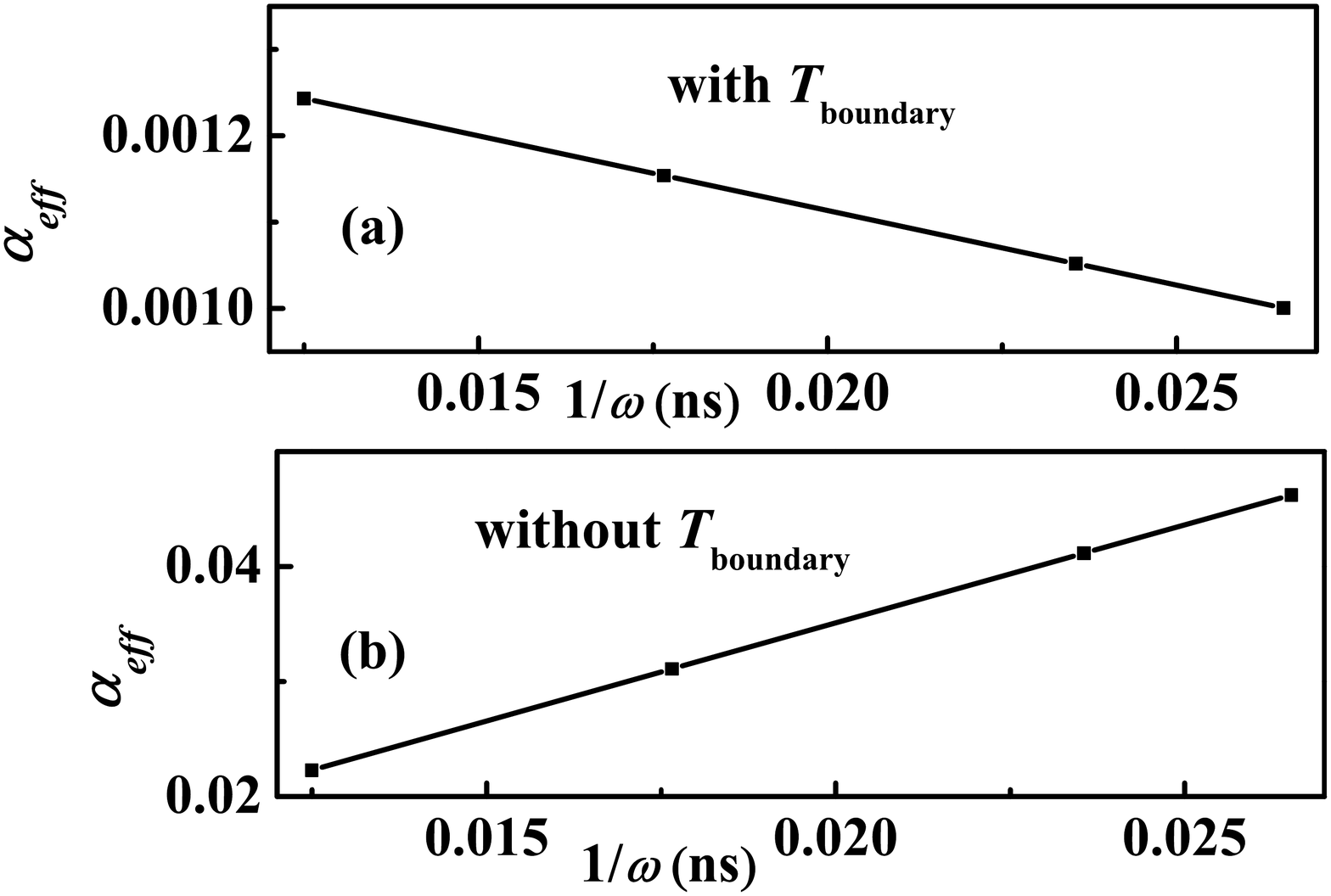}
    \caption{\label{alpha-omega} Influence of the inhomogeneous electric torque (IET) on the effective damping constant $ \alpha_{\rm {eff}} $ with (a) and without (b) the boundary torque ($ T_{\mathrm{boundary}} $),  plotted as a function of inverse frequency $ 1/\omega $. Here, $ \omega $ is the oscillation frequency (only four resonance frequencies are included). The electric field gradient is $ E_{gr} = 0.24 $ (MV/m)/nm.}
\end{figure}

\begin{figure}
    \includegraphics[width=0.48\textwidth]{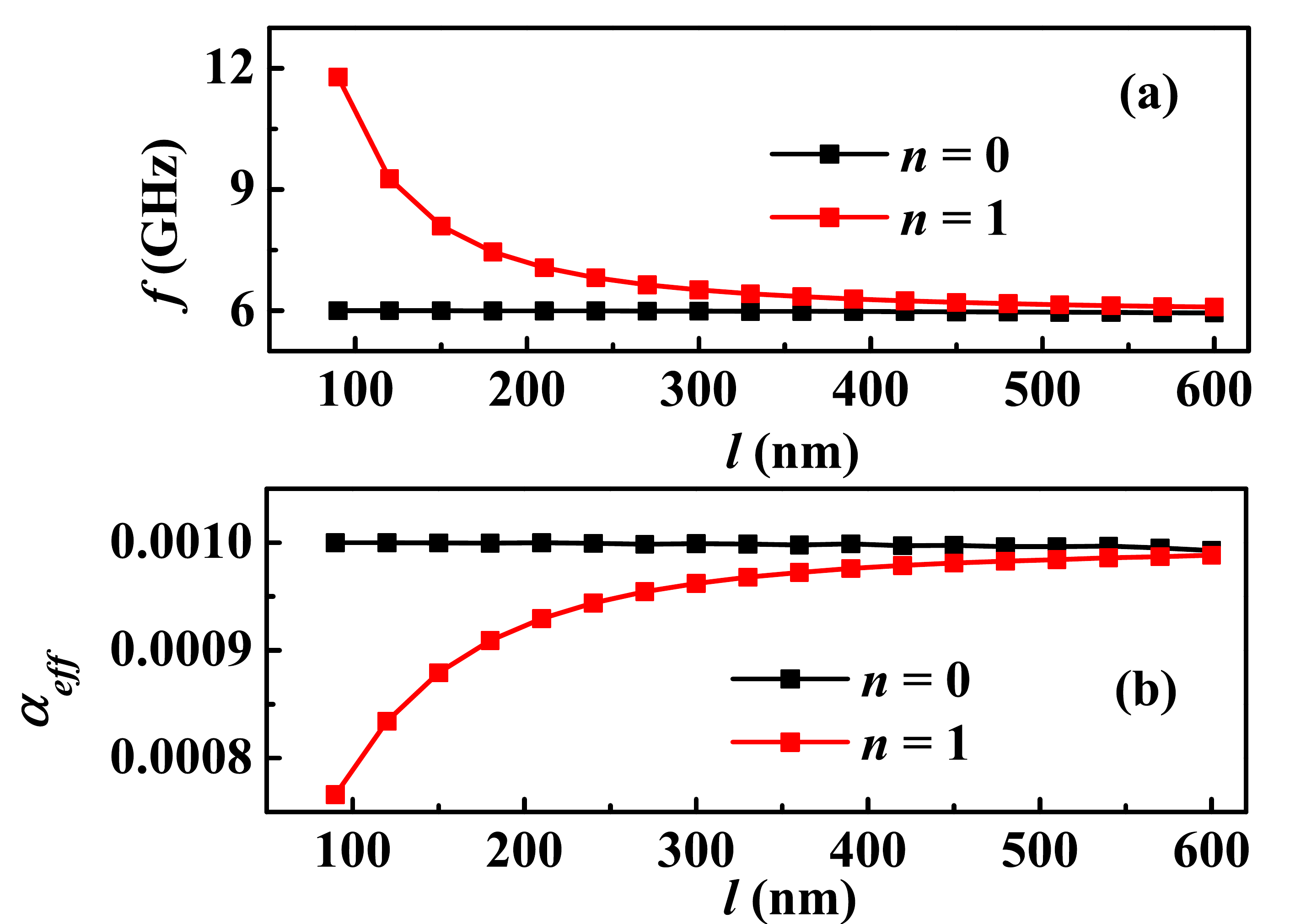}
    \caption{\label{alpha-width} The size effects for the oscillation frequencies and the effective damping are shown in (a) and (b).  For $ n = 0 $ and $ n = 1 $ modes,
the length of the sample $ l $ is varied between $100<x <600$ nm. The gradient of the electric field  is $ E_{gr} = -0.24 $ (MV/m)/nm.}
\end{figure}

As distinct from the current-induced Slonczewski torque, the ME coupling effects the magnetization precessional damping not only through the magnetoelectric torque term, i.e.,  Eq. (\ref{IET2}) (the torque related to the $ \partial_i \vec{m} $  in Eq. (\ref{IET1})  is rather weak for the resonant mode). The ME coupling impacts the boundary constraint and indirectly influences the effective damping.
We remove the IET part and write out the ME coupling induced magnetic torque under the boundary constraint (Eq. (\ref{boundary})):
\begin{equation}
\begin{aligned}
\displaystyle  \vec{T}_{(nx = 1)} &= -\frac{\gamma c_E E_{(nx = 1)}}{\mu_0 M_s \triangle_x } \vec{m}_{(nx = 1)} \times (\vec{m}_{(nx = 2)} \times \vec{p_E}), \\
\vec{T}_{(nx=n_0)} &= \frac{\gamma c_E E_{(nx=n)}}{\mu_0 M_s \triangle_x } \vec{m}_{(nx=n_0)} \times (\vec{m}_{(nx=n_0-1)} \times \vec{p_E}).\\
\end{aligned}
\label{boundary-field}
\end{equation}
The equation (\ref{boundary-field}) is expressed in the central derivative form and $ nx = 1 $ and $ nx = n_0 $ represent the cells located at the left ($ x = -125 $ nm) and right ($ x = 125 $ nm) boundaries, respectively. Besides the IET (Eq. (\ref{IET2})), obviously there is another additional damping-like torque associated with the effect of boundaries. This additional damping-like torque also modifies the magnetization dynamics.
The effective damping $ \alpha_{\mathrm{eff}} $ is determined by the context of Eqs.(\ref{IET2}) and (\ref{boundary-field}). For the electric field profile $ \vec{E} = (0,E_{gr} x \vec{e}_y,0) $, adopted in our simulations, the directions of boundary and IET torques in Eqs.(\ref{IET2}) and (\ref{boundary-field}) are always opposite.
For the $ n = 0 $ resonance mode, two neighboring spins are always parallel ( $ \vec{m}_{(nx = 1)} \approx \vec{m}_{(nx = 2)} $ and $ \vec{m}_{(nx=n_0)} \approx \vec{m}_{(nx = n_0-1)} $ ) in Eq. (\ref{boundary-field}). The Fig. \ref{alpha} demonstrates that for the $ n = 0 $ resonance mode, two opposite torques (boundary torque and IET) totally compensate each other. For higher standing wave nodes and frequencies, the oscillation of non-collinearity between two neighboring spins increases. This relaxes the boundary torque in Eq. (\ref{boundary-field}) and as a consequence the boundary torque does not compensate IET any more.
Therefore, the change in $ \alpha_{eff} $ is enhanced for higher resonance modes. We plot the variation of the effective damping constant $ \alpha_{eff} $ in the presence and absence of the boundary torque, see Fig. \ref{alpha} and Fig. \ref{alpha-STT} respectively. Variation of the effective damping constant $ \alpha_{eff} $ in the absence of the boundary torques Fig. \ref{alpha-STT} is caused solely by IET. Comparing Fig. \ref{alpha} and Fig. \ref{alpha-STT} we see that the boundary torque reduces the variation of the effective damping constant $ \alpha_{eff} $.
It is instructive to explore the dependence of the effective damping constant $ \alpha_{eff} $ on the inverse frequency $ 1/\omega $. The result is presented in Fig. \ref{alpha-omega}.
Evidently, the effective damping constant $ \alpha_{eff} $  depends linearly on the inverse frequency. This result is similar to the case of the current-induced spin-transfer torque \cite{Krivorotov2005}.
The wavelength and the frequency of the oscillation mode can be steered with the magnet's length $ l $. This also changes the effective damping.
As is shown in Fig. \ref{alpha-width}(a), with the increase of $ l $, the oscillation frequency of $ n = 1 $ mode decreases. Similar to Fig. \ref{alpha-omega}(a), the decreasing of the frequency decreases the variation of the effective damping (Fig. \ref{alpha-width}(b)).

\begin{figure}
    \includegraphics[width=0.48\textwidth]{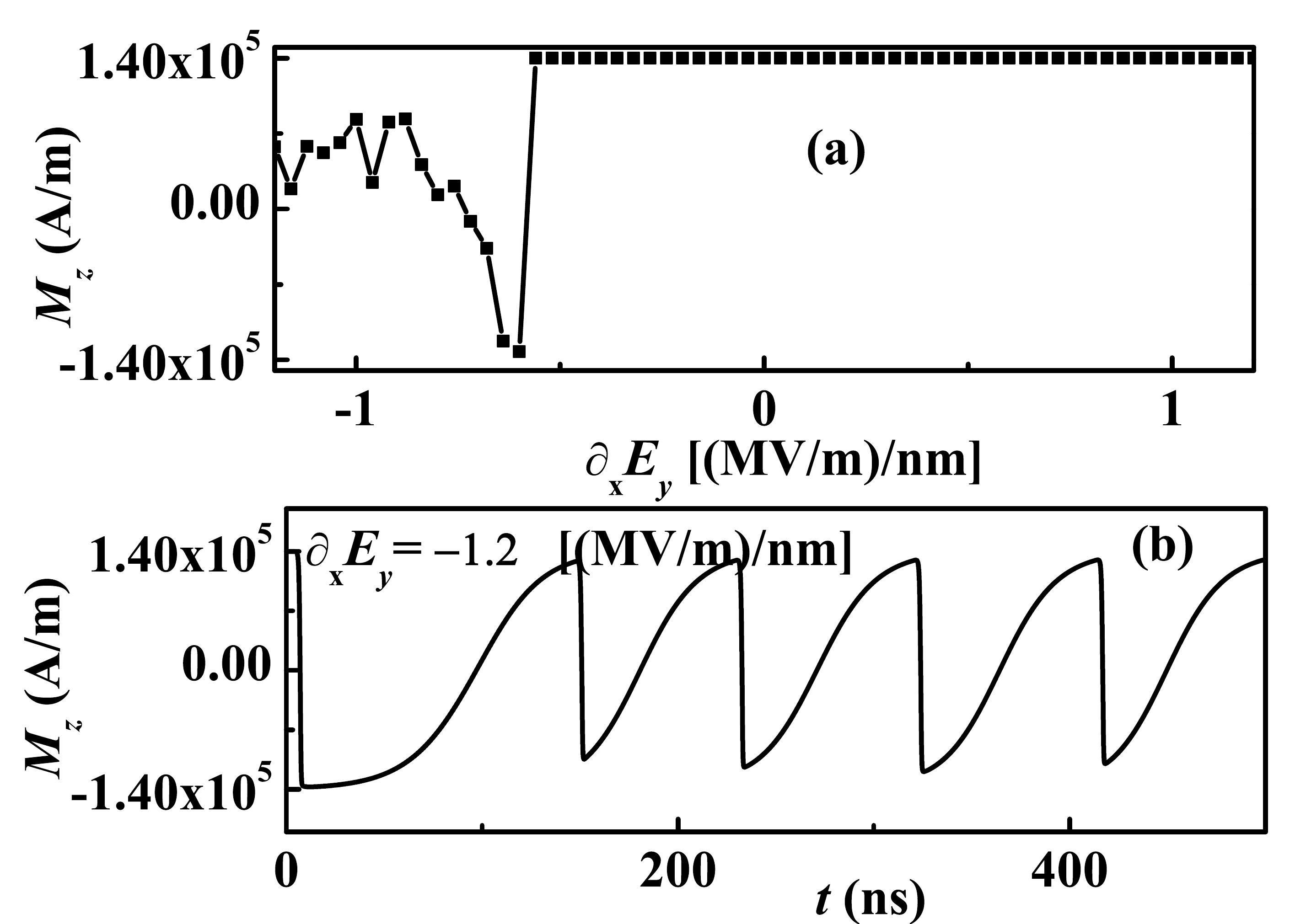}
    \caption{\label{auto-oscilla} (a) The averaged $ M_z $ after $ t > 90 $ ns, as a function of the inhomogeneous electric field $ \partial E_y $.  (b) The averaged  $ M_z $  as a function of time. The inhomogeneous electric field with the constant gradient $ \partial_xE_y = -1.2  $ (MV/m)/nm is applied along the $y$ axis. The magnetic field  $ H_z = 1.6 \times 10^5 $ A/m  is applied along the $ z $ axis, and the initial magnetization is aligned along the $ +z $ axis.}
\end{figure}

\begin{figure}
    \includegraphics[width=0.48\textwidth]{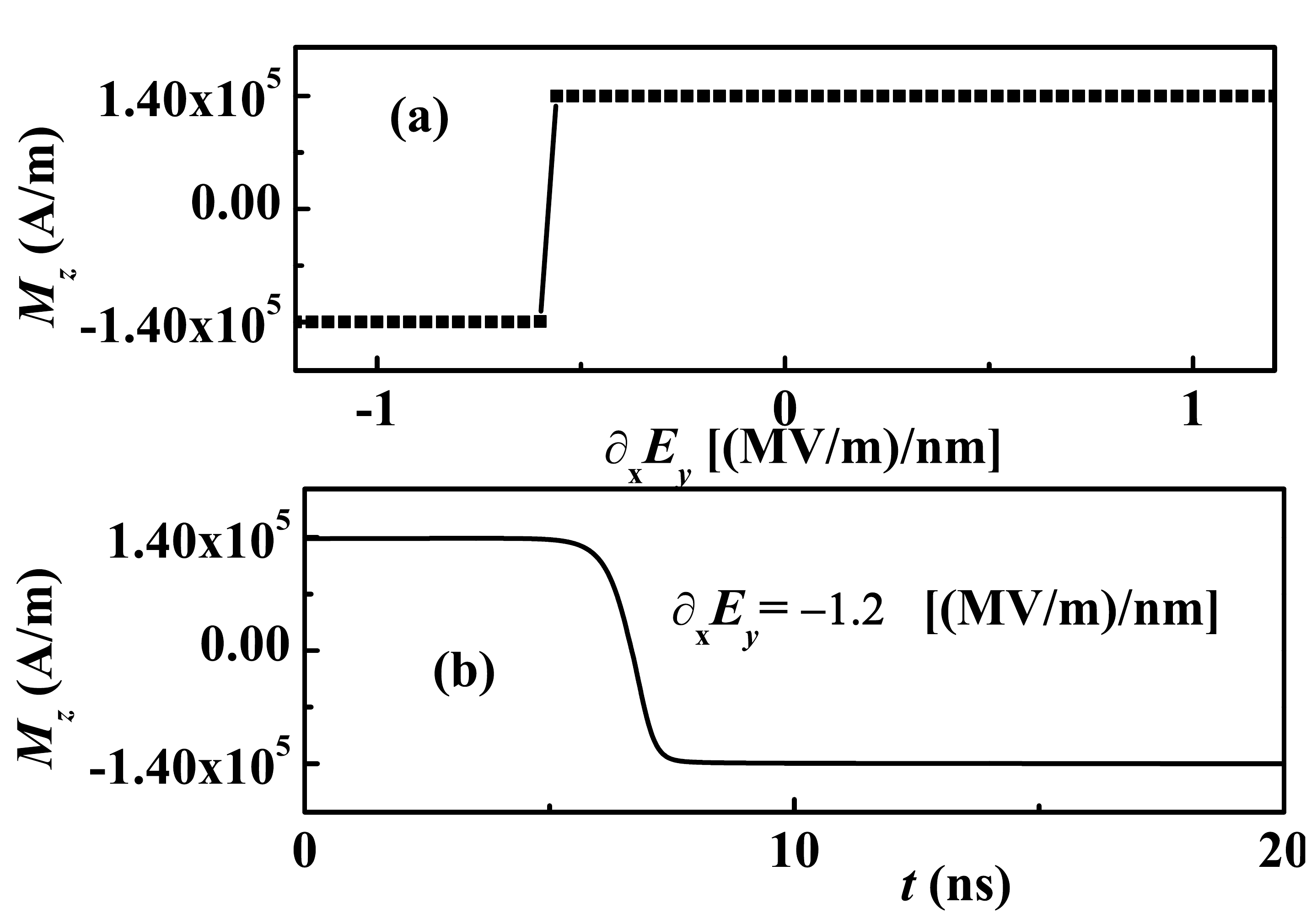}
    \caption{\label{reverse} (a) The averaged magnetization component $ M_z $  after $ t = 90 $ ns  as a function of the inhomogeneous electric field $ \partial_xE_y $.  (b) The time dependence of the averaged  magnetization component $ M_z $ . The  electric field gradient  is equal to $ \partial_xE_y = -1.2 $ (MV/m)/nm, the magnetocrystalline anisotropy field has the form $ \vec{H}_{k} = 2 K_z m_z / (\mu_0 M_s) \vec{e}_z $, where $ K_z= 2.3 \times 10^{4} $ J/m$ ^3 $,  and the initial magnetization is aligned along the $ +z $.}
\end{figure}

Furthermore, employing a large enough negative $ \partial_xE_y $, we successfully reverse the magnetization direction and drive the auto-oscillation. This fact testifies the feature of the spin transfer like torque. The averaged $ M_z $ as a function of  $ \partial_xE_y $ is shown in Fig. \ref{auto-oscilla}(a). The external magnetic field $ H_z = 1.6 \times 10^5 $ is applied along the $ z $ axis and the initial magnetization is along the $ +z $. When the negative $ \partial_xE_y $ is large enough, the $+z$ local magnetization becomes unstable after 90 ns of evolution. For $ \partial_xE_y = -1.2 $ (MV/m)/nm, the averaged  $ M_z $ reveals the nature of  magnetization oscillation induced by IET, as demonstrated in Fig. \ref{auto-oscilla}(b). Besides, by setting an anisotropy field  $ \vec{H}_{k} = 2 K_z m_z / (\mu_0 M_s) \vec{e}_z $ with $ K_z= 2.3 \times 10^{4} $ J/m$ ^3 $ instead of $ H_z $, the direction of the equilibrium magnetization can be reversed by IET, as is shown in Fig. \ref{reverse}, and there is no auto-oscillation in this case.

\section{thermal magnonic spin current}
\label{sec:spincurrent}

Conjointly with the uniform temperature bias,  STT induces a non-equilibrium magnon flow, meaning a thermally assisted magnonic spin current \cite{wang2017}. The thermal random magnetic field activates the magnetization oscillations around the equilibrium state and excites the thermal magnons.
In this way, the electric spin-polarized current can be converted into a magnonic current employing costly procedure. To assess the costs, one should include ohmic losses in the generation of the electric spin-polarized current and estimate the current conversion efficiency. In the present work, we show that the IET can be used to generate non-equilibrium magnonic flow. The underlying physical mechanism of our method is based on an inhomogeneous electric field $ E_y $  that modifies the thermal magnon density profile.

\begin{figure}
    \includegraphics[width=0.48\textwidth]{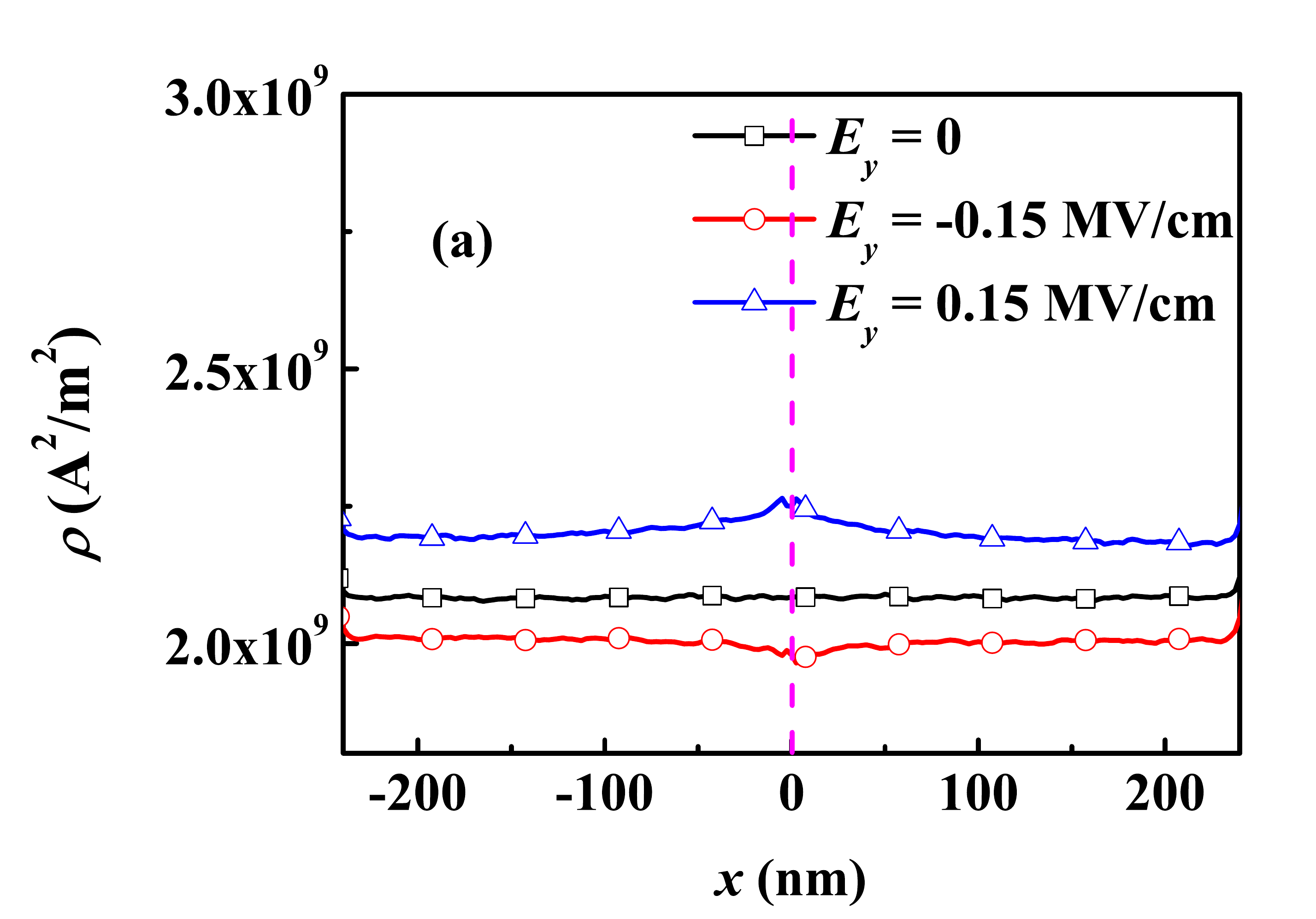}
    \includegraphics[width=0.48\textwidth]{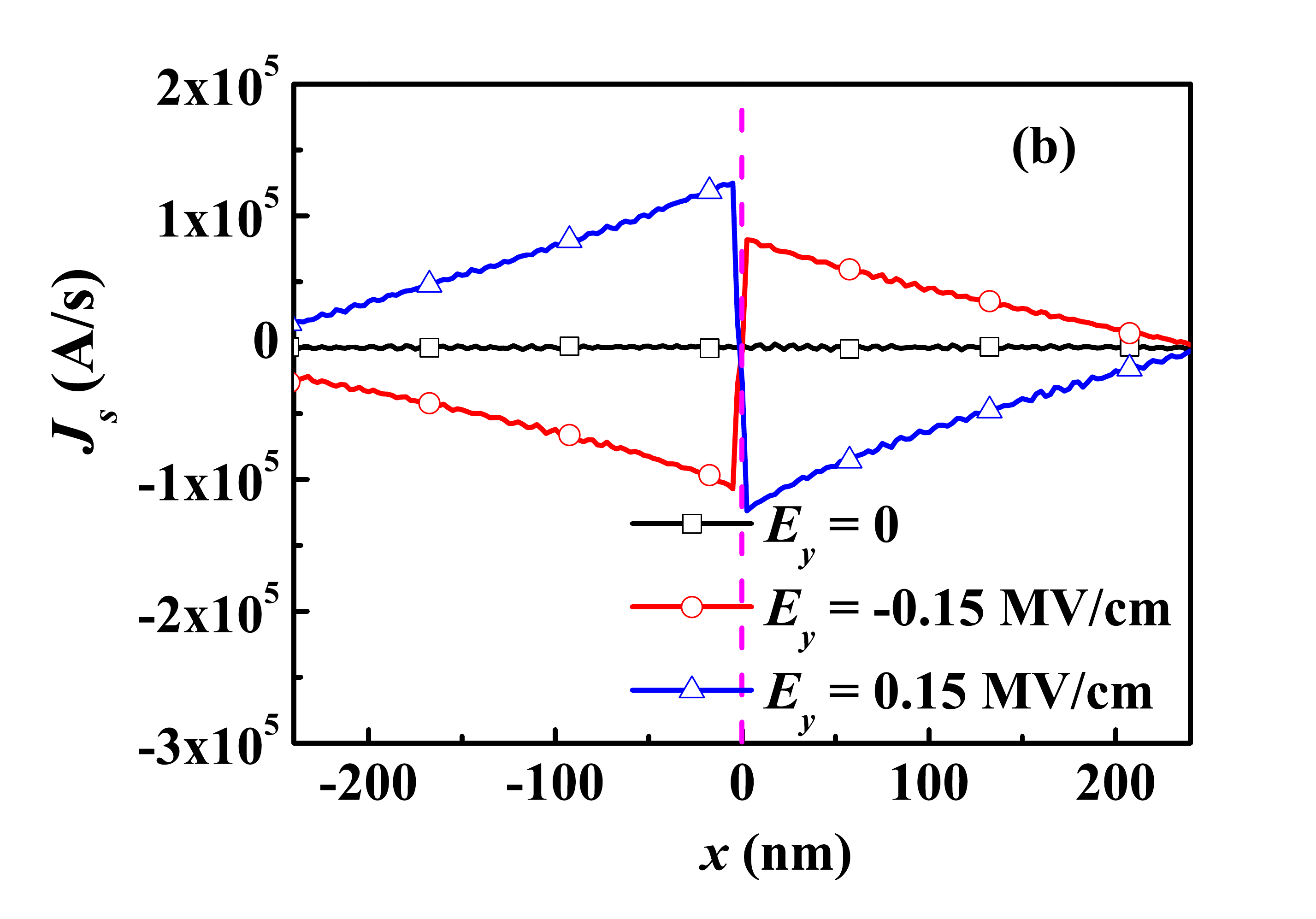}
    \caption{\label{spin current} Profiles of the averaged magnon density $ \rho $ (a) and the magnonic spin current $ J_{s} $ (b) flowing along the $ x $ axis. The electric field $ E_y = \pm 0.15 \mathrm{MV/cm} $ is applied in the left part of the sample ( $ x < x_0 = 0 $ ). }
\end{figure}

\begin{figure}
    \includegraphics[width=0.48\textwidth]{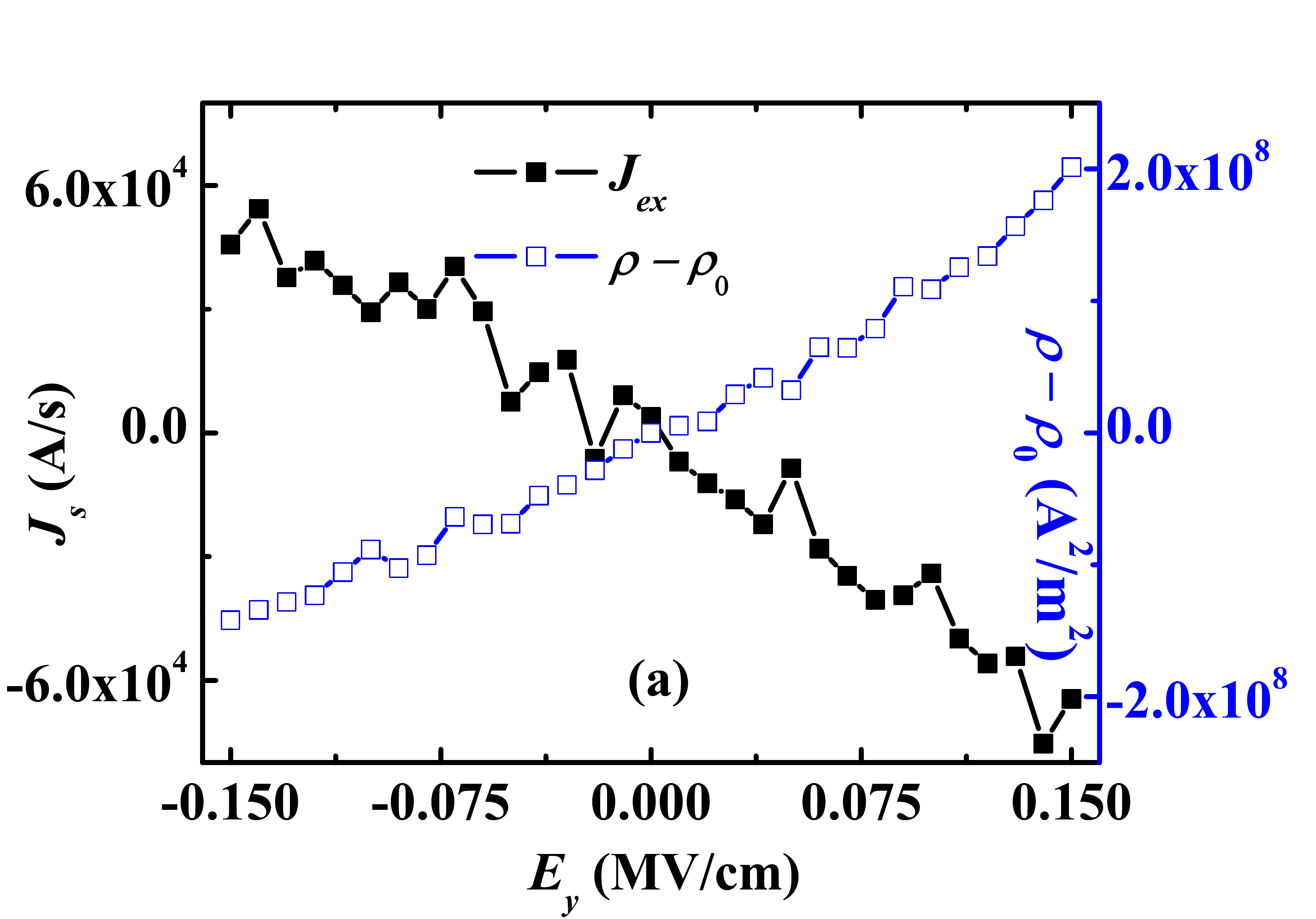}
    \includegraphics[width=0.48\textwidth]{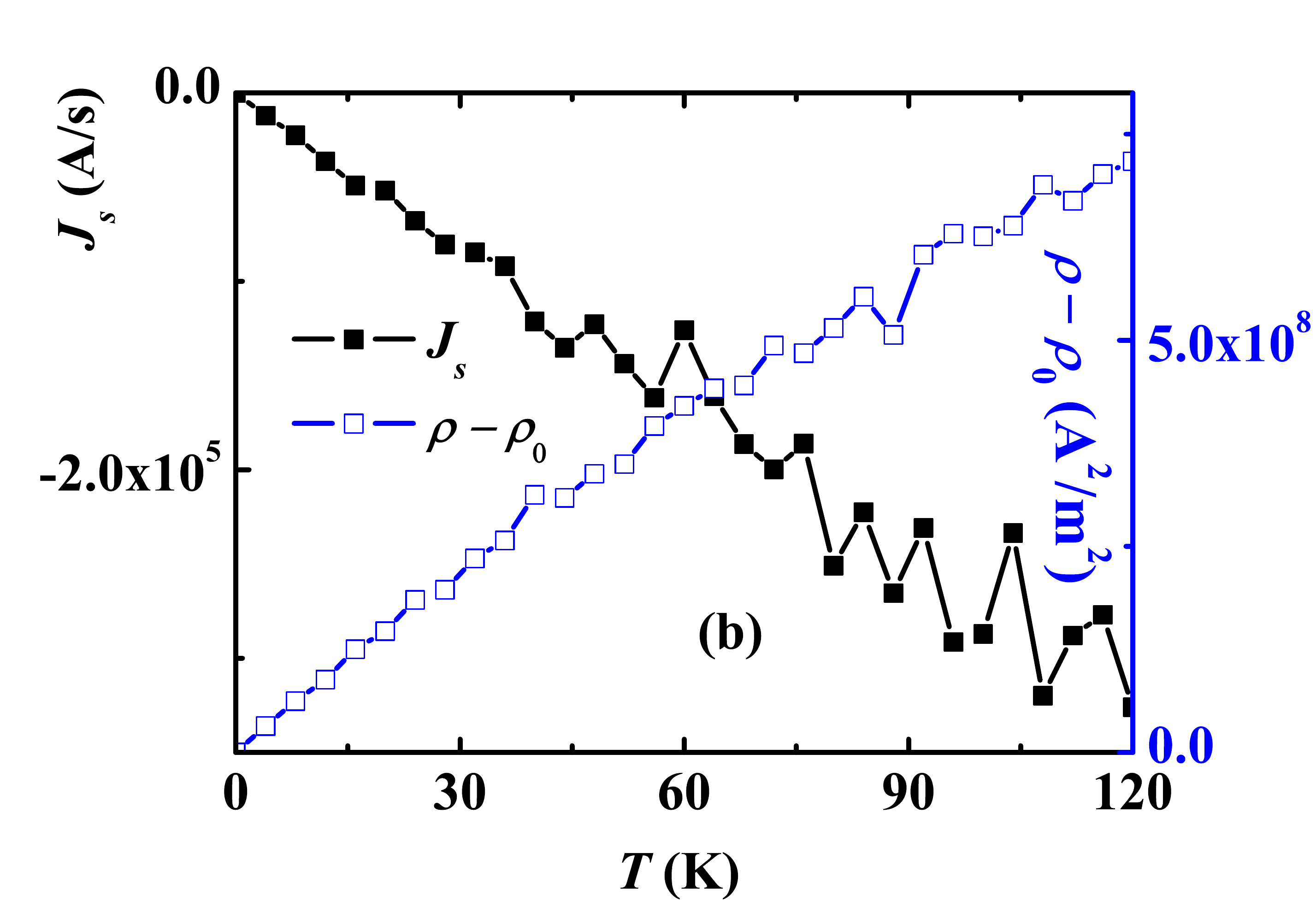}
    \caption{\label{current-E-T} Averaged magnon density $ \rho - \rho_0 $ and magnonic spin current $ J_{s} $ as a function of electric field $ E_y $ (a) and temperature $ T $ (b). $ T = 25 K $ for (a) and $ E_y = -0.15 $ MV/cm for (b).  $ \rho_0 $ represents the magnon density in absence of the electric field ($ E_y = 0 $). }
\end{figure}
Fig. \ref{spin current} shows the spatial distribution of the magnon density $ \rho $ and the magnonic spin current $ J_{ex} $. The temperature $ T $ is taken to be 25 K, the component of the electric field  $ E_y = \pm 0.15 \mathrm{MV/cm} $ is  applied only in the region $ x < x_0 = 0 $. The magnon density $ \rho = M_x^2 + M_y^2 $ is quantified by the squared transversal magnetization components and the total magnonic spin current is equal to  $ J_{s} = J_{ex} +J_{E}$, where $ J_{ex} = \frac{2 \gamma A_{ex}}{\mu_0 M_s^2} (M_x \partial_x M_y - M_y \partial_x M_x)  $ is the exchange spin current and the chiral spin current is defined as follows $ J_{E} = -\frac{\gamma c_E E_y}{\mu_0 M_s^2} (M_x^2 + M_y^2) $ see \cite{Wang2016}. In the region ($ x > x_0 $) without the electric field, the chiral spin current is zero and the total current is equal to the exchange spin current $ J_{s} = J_{ex} $.
In the absence of the electric field ($ E_y = 0 $), the uniform temperature cannot induce a net  magnonic spin current. When the electric field $ E_y = 0.15 $ MV/cm is applied in the region $ x < x_0 $, the density of thermal magnons  is obviously enhanced at the interface $ x = x_0 $. Meanwhile, non-equilibrium magnons diffuse away from the $ x = x_0 $ and generate the negative (positive) magnonic spin current $ J_{s} $ in the right (left) side. $ J_{s} $ decreases with $ x $ due to the attenuation and boundary reflection. Reversal of the electric field ($ E_y = -0.15 $ MV/cm) reduces the thermal magnons at $ x = x_0 $, and $ J_{s} $ in the right (left) side becomes positive (negative). Besides, the effect of the enhancement of the thermal magnons ($ E_y > 0 $) is obviously stronger than the reduction effect ($ E_y < 0 $), as shown in Fig. \ref{current-E-T}(a). The variation of the magnon density $ \left| \rho - \rho_0 \right| $ and the  magnonic spin current $\left| J_{s} \right|$ both increases with the increase of $ \left| E_y \right| $. Here, $ \rho_0 $ represents the magnon density without electric field ($ E_y = 0 $).  Moreover, an increase in the uniform temperature $T$ enhances the variation of the magnon density and the magnonic spin current $ J_{s} $, as shown in Fig. \ref{current-E-T}(b).

\section{Skyrmion motion driven by inhomogeneous electric field}
\label{sec:motion}

\begin{figure}
	\includegraphics[width=0.48\textwidth]{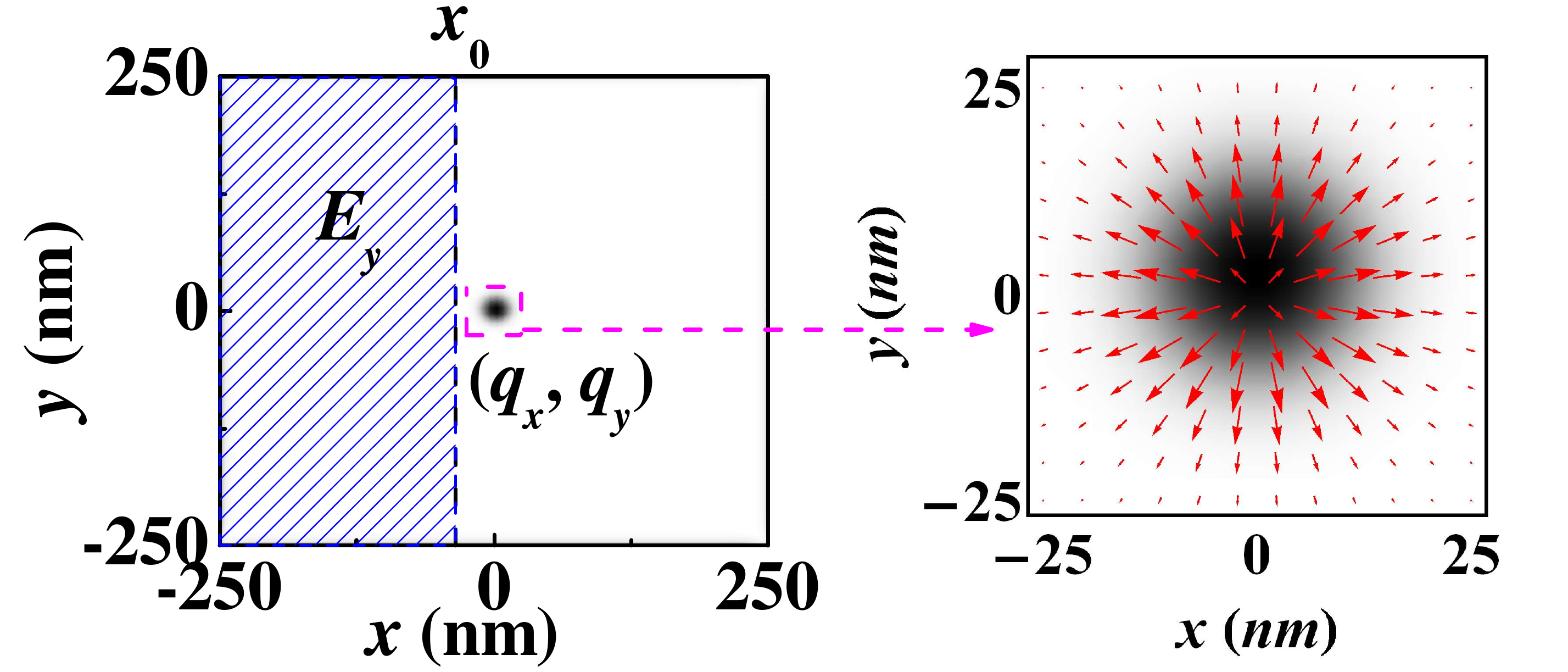}
	\caption{\label{Fig1} Skyrmion motion induced by the IET and by the thermal magnonic spin current.  Initially, the skyrmion is located in the center ($ q_x = 0 $ and $ q_y = 0 $) of the sample. The temperature $ T $ is uniform in the whole sample. The electric field  $ E_y $ is applied in the left part of the sample ($ x < x_0 $) and  generates the magnonic current flowing  from (or toward) the interface ($ x = x_0 $). The magnonic current drives the skyrmion.}
\end{figure}

\begin{figure}
	\includegraphics[width=0.48\textwidth]{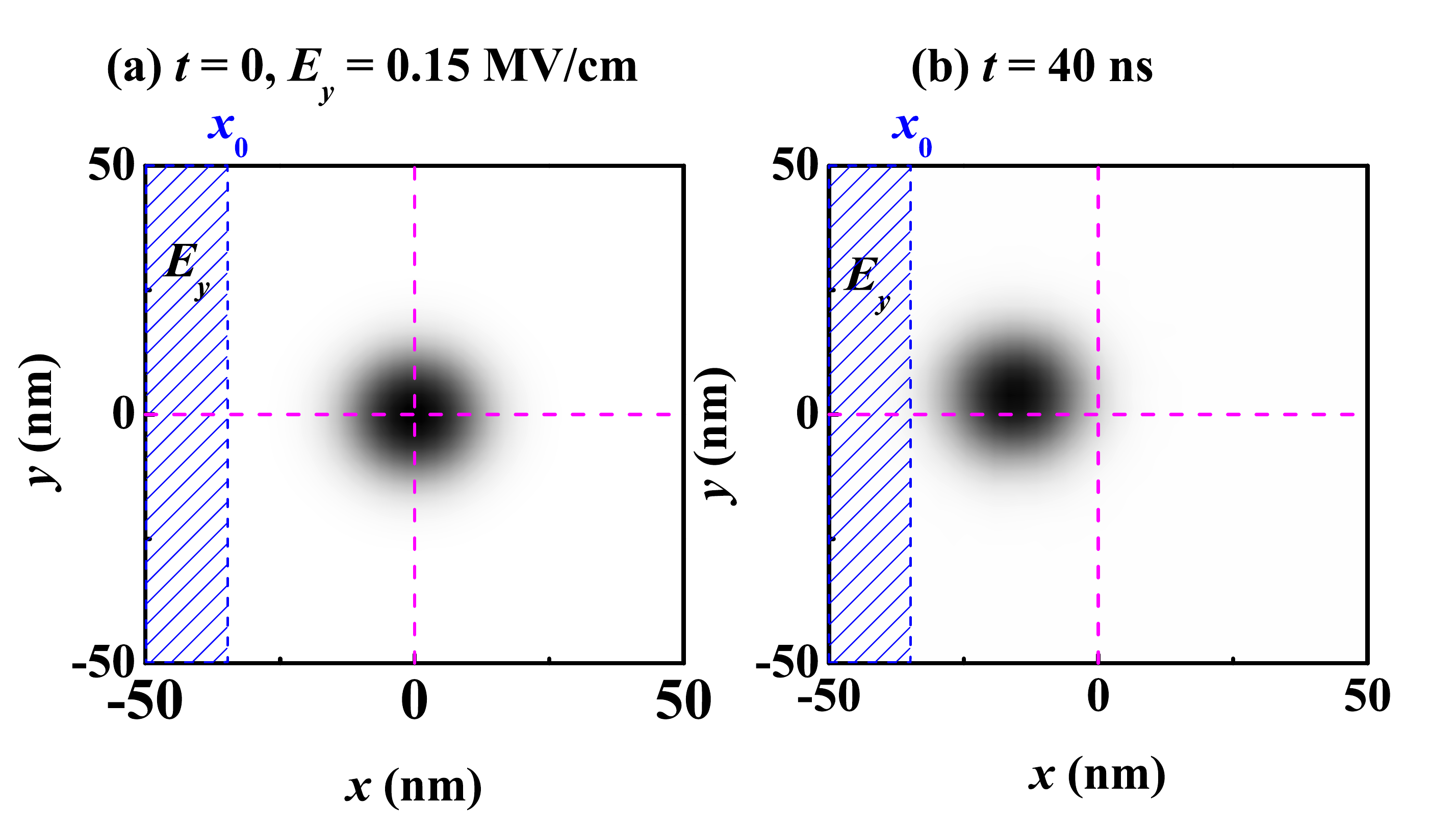}
	\caption{\label{skyrmion-motion2} Snapshots of Skyrmion motions taken in the different moments of time. Initally ((a) $ t = 0 $), the skyrmion is located at $ q_x = 0 $ and $ q_y = 0 $. The applied electric field $ E_y = 0.15 \mathrm{MV/cm} $  and the uniform temperature $ T = 25 $ K,  moves the skyrmion toward the interface $ x=x_0 =-35 $ nm.}
\end{figure}

\begin{figure}
    \includegraphics[width=0.48\textwidth]{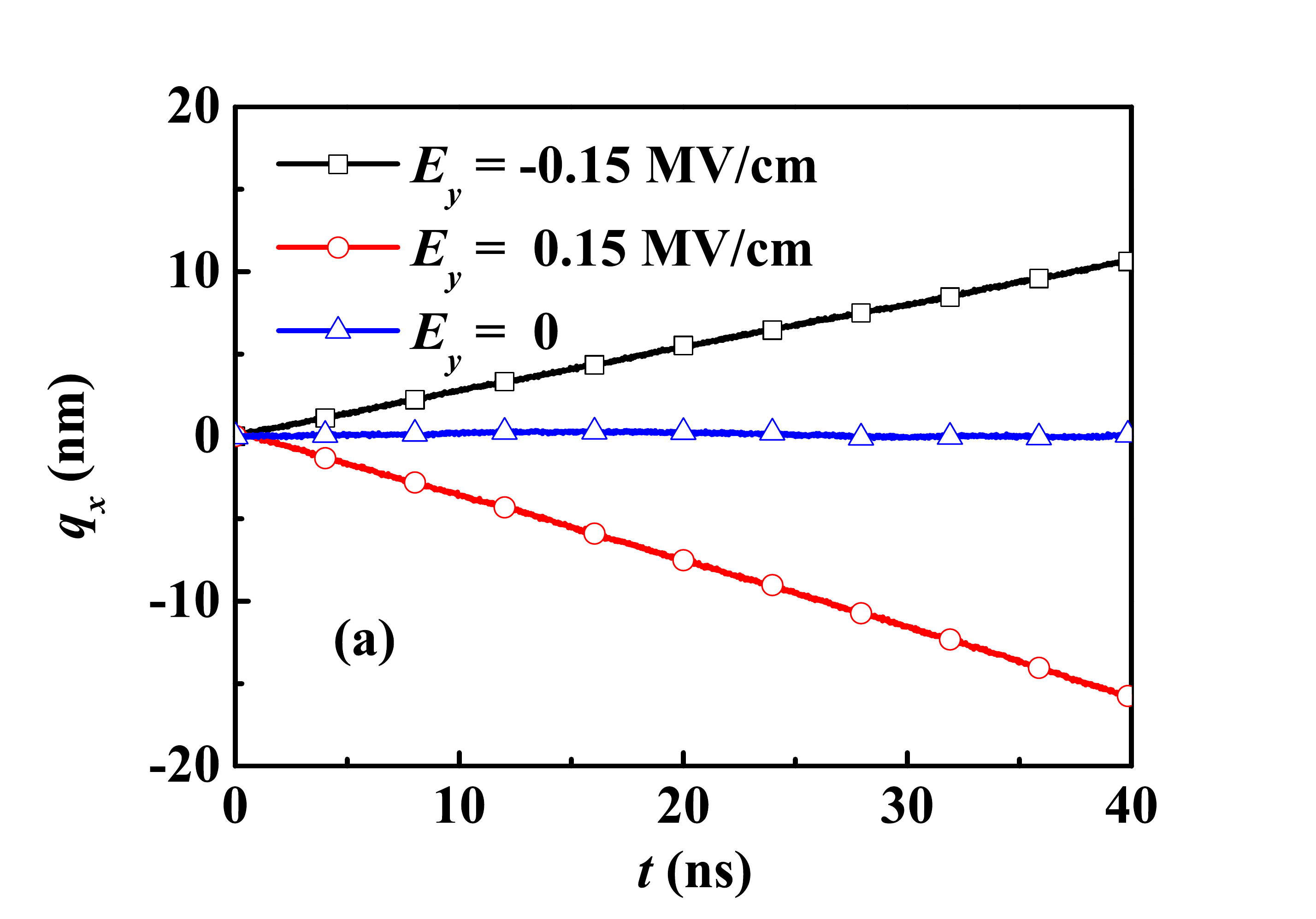}
    \includegraphics[width=0.48\textwidth]{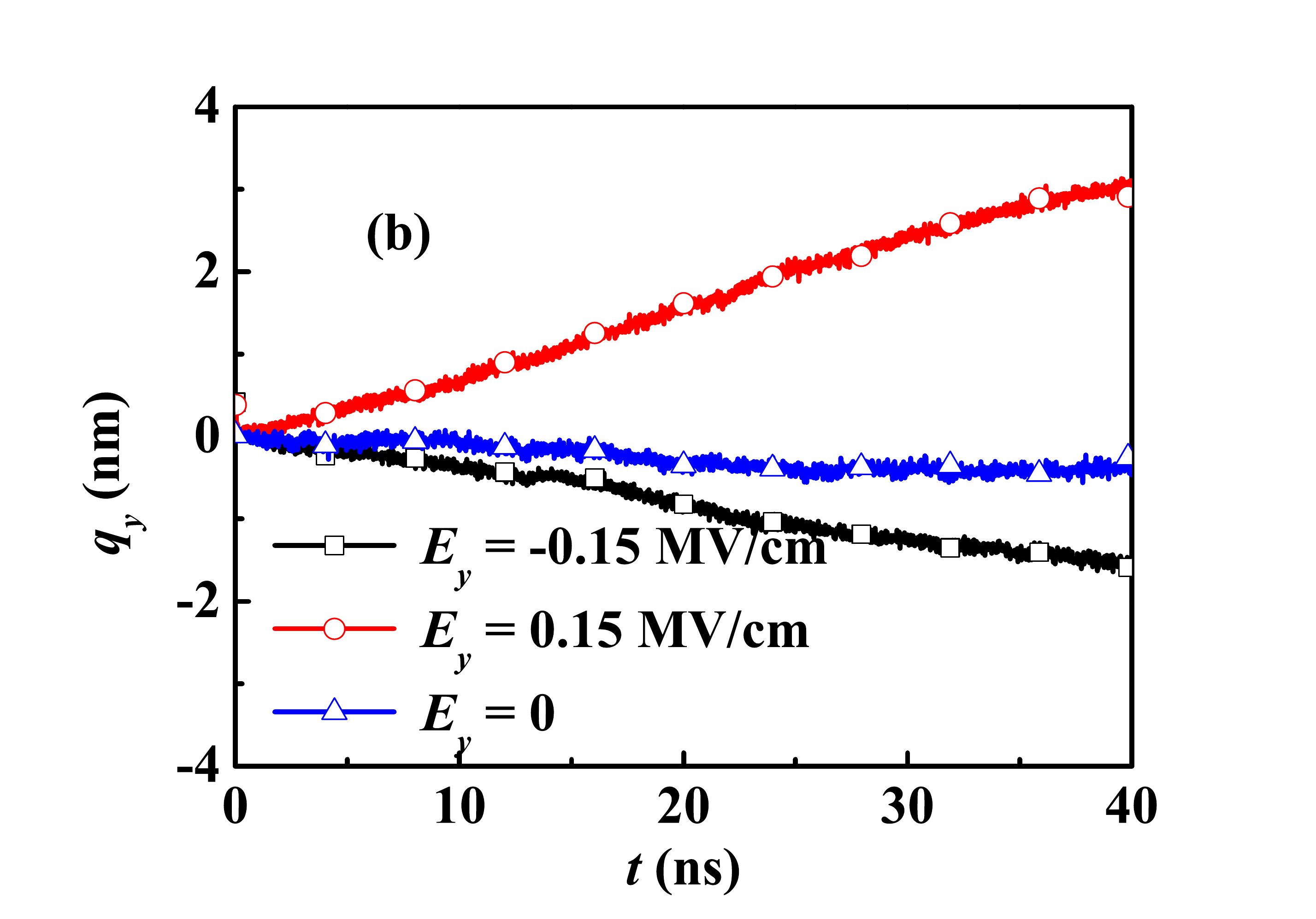}
    \caption{\label{skyrmion motion}  Motion of the skyrmion center   ($ q_x, q_y $) in time, in the $ x $ (a) and $ y $ (b) directions. Electric fields with $ E_y = \pm 0.15 \mathrm{MV/cm} $ are  applied in the left part ( $ x < x_0 = -35 $ nm ). Temperature is equal to $ T = 25K $.}
\end{figure}

\begin{figure}
    \includegraphics[width=0.48\textwidth]{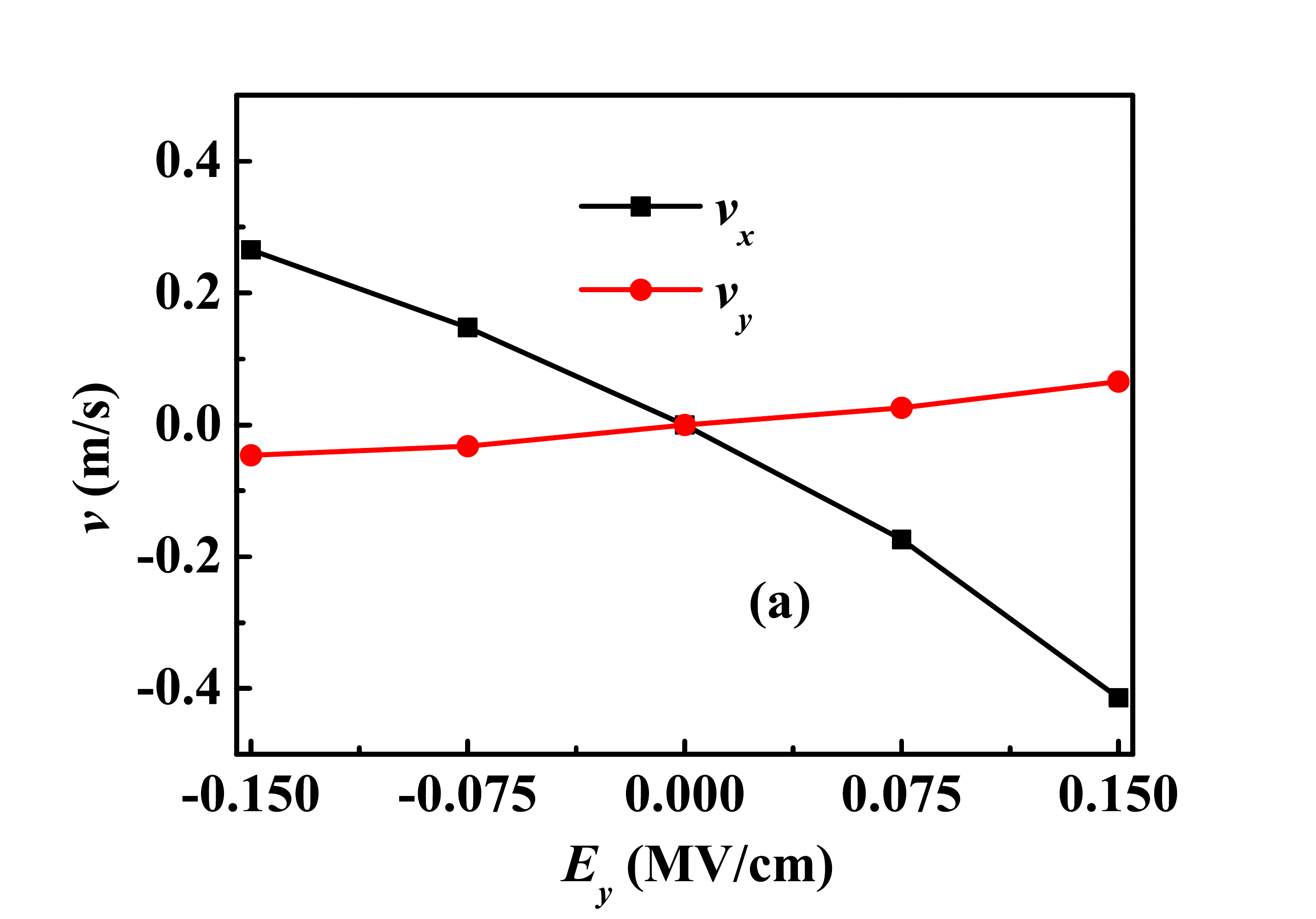}
    \includegraphics[width=0.48\textwidth]{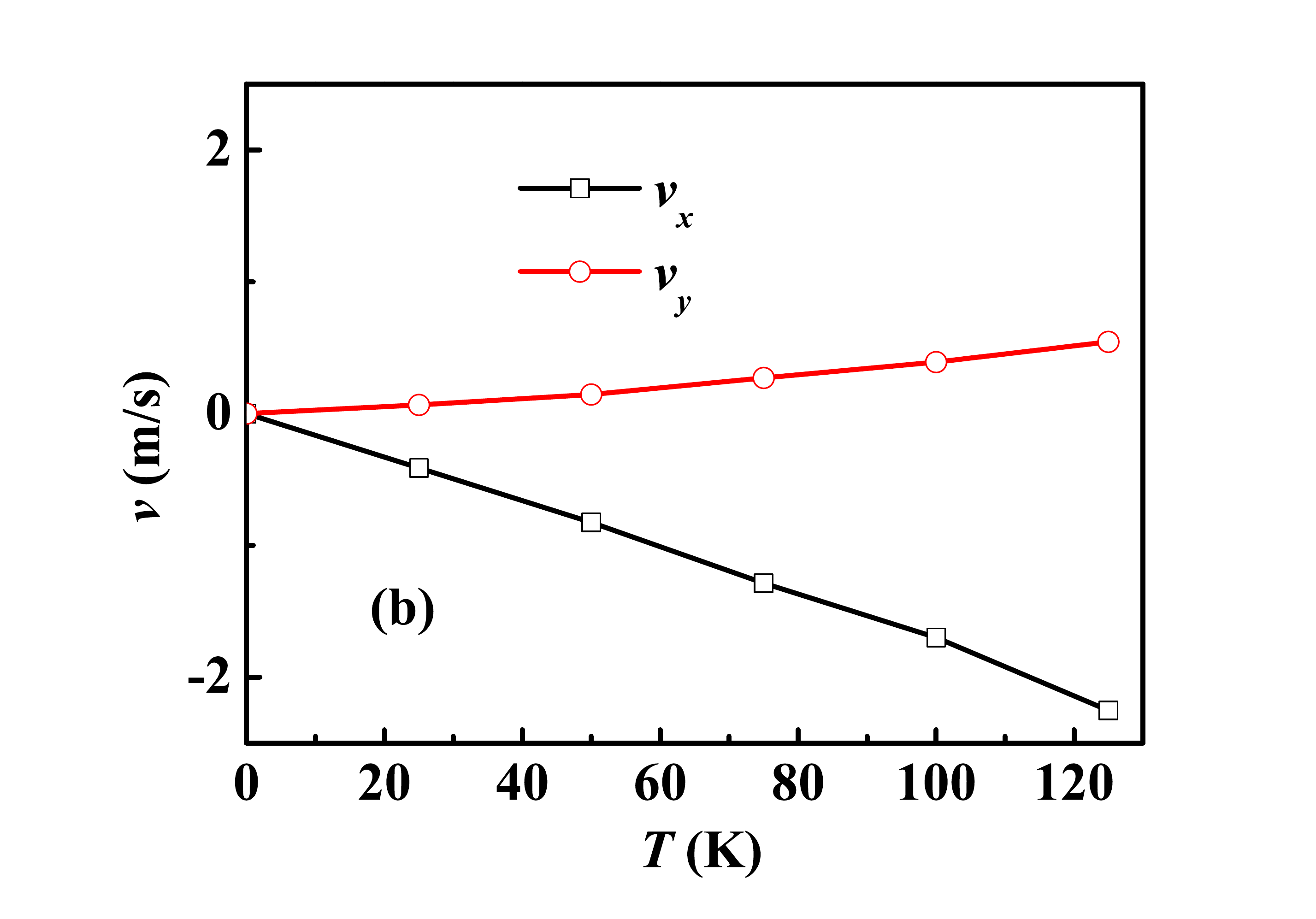}
    \caption{\label{speed-T-ey}  Skyrmion velocities along the x axis $ v_x $ and  y axis $ v_y $ as functions of the electric field $ E_y $ for $ T = 25 K $ (a) and the temperature $ T $  for $ E_y = -0.15 $ MV/cm (b).}
\end{figure}

To contrast our inhomogeneous E-field-based method  with standard recipes of skyrmion motion based either on the electric spin torque current or on thermally assisted magnonic current, we embed the skyrmion at the right side of the interface $ x = x_0 = -35 $ nm and let the induced thermal magnonic spin current $ J_{s} $  drives it (as demonstrated by Fig. \ref{Fig1}).
As is shown in Figs. \ref{skyrmion-motion2} and \ref{skyrmion motion}, the skyrmion moves towards the interface $ x = x_0 $ along the $ x $ axis when the thermal magnon density is enhanced and the exchange magnon spin current is negative $ J_{ex} < 0 $ for $ E_y > 0 $.  Besides,  the skyrmion drifts along the $ y $ axis due to the skyrmion Hall effect. Switching sign of the electric field ($ E_y < 0 )$ leads to an inversion of the skyrmion motion in both $x$ and $y$ directions (Fig. \ref{speed-T-ey}(a)), however with smaller velocities. Besides, the  motion at the higher temperature is faster (Fig. \ref{speed-T-ey}(b)), i.e. thermal effect enhances the skyrmion velocity.

\begin{figure}
    \includegraphics[width=0.48\textwidth]{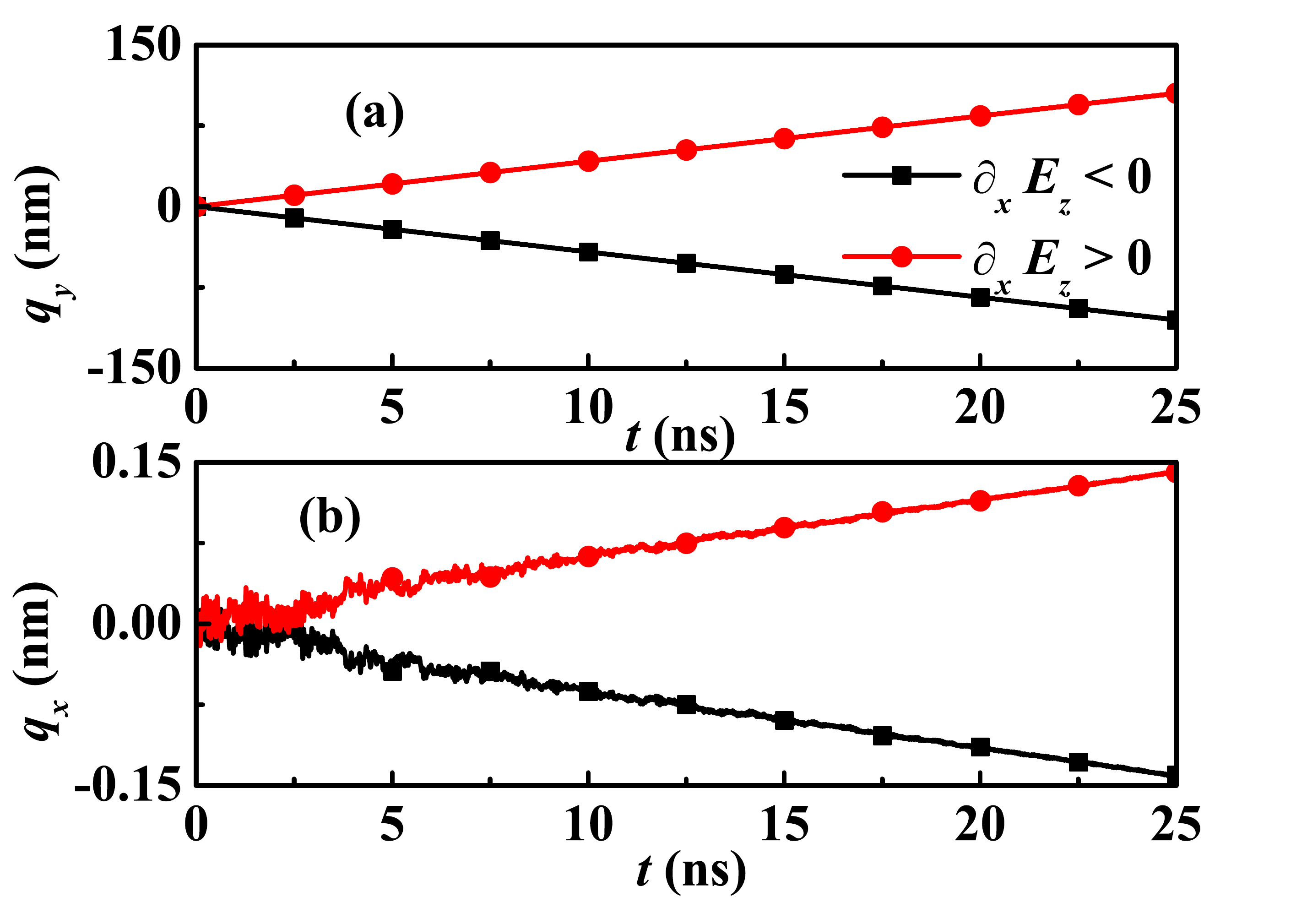}
    \caption{\label{ez-skyrmion-time}  Motion of the center of skyrmion ($ q_x, q_y $) in time, in the $ x $ (b) and $ y $ (a) directions. The electric field with $ E_z = \pm E_{gr} x $,  $ E_{gr} = 0.47 $ (MV/m)/nm is applied. The temperature is equal to $ T = 0 $.}
\end{figure}

The STT with spin polarization $ \vec{p} $ flows along the $ y $ axis $ c_j > 0 $ and  drags  the skyrmion in  both  $ x $ and $ y $ directions simultaneously at zero temperature $T=0$.  The corresponding velocities $ v_x $ and $ v_y $ satisfy the condition $ v_x = \alpha v_y $ \cite{yzhou2016}. Due to the equivalence of IET and the STT, an interesting question is whether IET as well can drag skyrmion at zero temperature, without thermal assistance and for the same electric field. The answer to this question is positive, as inferred from Fig. \ref{ez-skyrmion-time}.

\begin{figure}
    \includegraphics[width=0.48\textwidth]{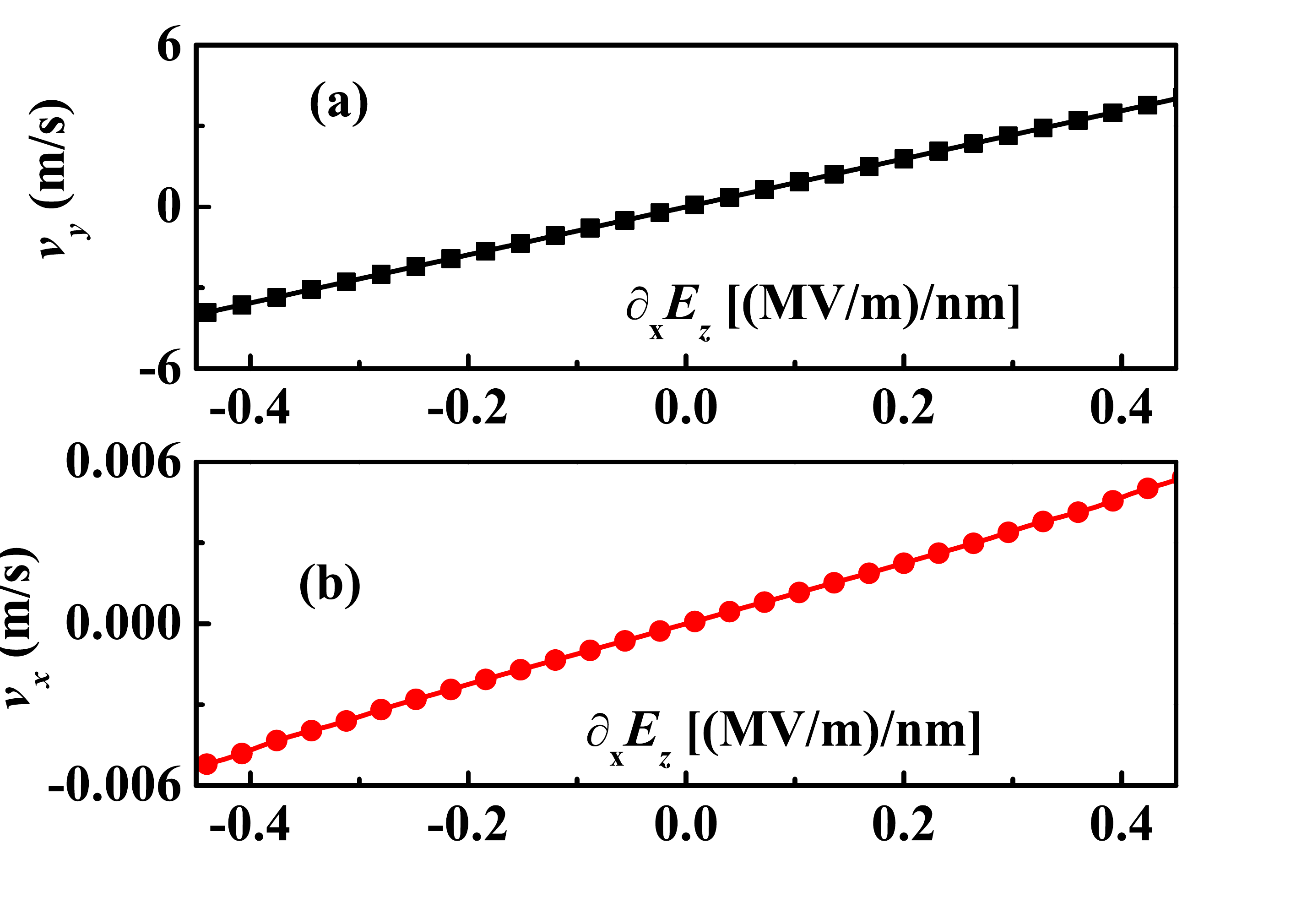}
    \caption{\label{vxy-ez}  Skyrmion velocities along the y axis $ v_y $ (a) and x axis $ v_x $ (b) as a function of the electric field gradient $ \partial_x E_z $.}
\end{figure}

Fig. \ref{ez-skyrmion-time}  illustrates the motion of the skyrmion driven by an electric field with a uniform gradient  $ \vec{E} = E_{gr} x \vec{e}_z $ MV/m,  $ \vec{p}_E = -\vec{e}_y $ at  zero temperature $ T = 0 $, for $ E_{gr} = 4.7 \times 10^{14} > 0 $. The skyrmion moves in the $ +y $ direction with the velocity $ v_y = 4.2 $ m/s. Due to the skyrmion Hall effect, the skyrmion is drifted along the $ +x $, and the velocity is $ v_x = 0.005 \mathrm{m/s} \approx \alpha v_y  $ \cite{yzhou2016}. Reversing $ E_{gr} $ leads to an inversion of the skyrmion motion. Besides, the skyrmion speed increases linearly with $ E_{gr} = \partial_x E_z $, as demonstrated by Fig. \ref{vxy-ez}. In this scenario the spatially homogeneous electric field cannot move the skyrmion.

\begin{figure}
    \includegraphics[width=0.48\textwidth]{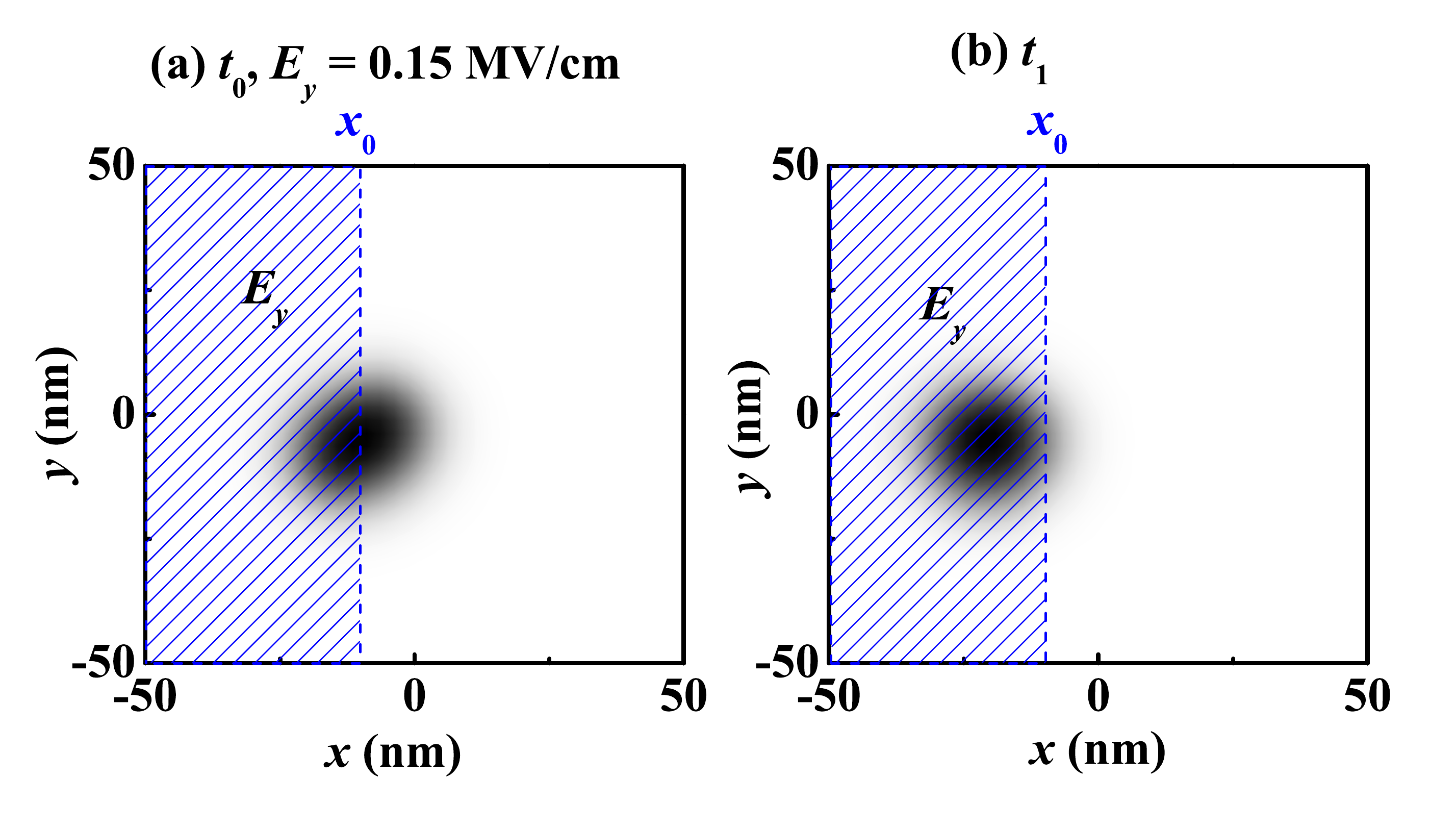}
    \caption{\label{Snapshots}  Snapshots of the skyrmion crossing the interface ($x = x_{0}$). Initially, the skyrmion is located at $q_{x} = 0$, $q_{y} = 0$ and driven by an external force, the skyrmion crosses the interface.}
\end{figure}

The important issue is the stability of the skyrmion when passing the interface, i.e. the region of the large electric field gradient. The strong inhomogeneity of the electric field may have an impact on the stability of the skyrmion.   The  calculations in  Fig. \ref{Snapshots}  demonstrate the stability of the skyrmions 
	 for the considered electric field parameters  when traversing the interface.

\section{Time-dependent electric field}
\label{pulse}

\begin{figure}
    \includegraphics[width=0.48\textwidth]{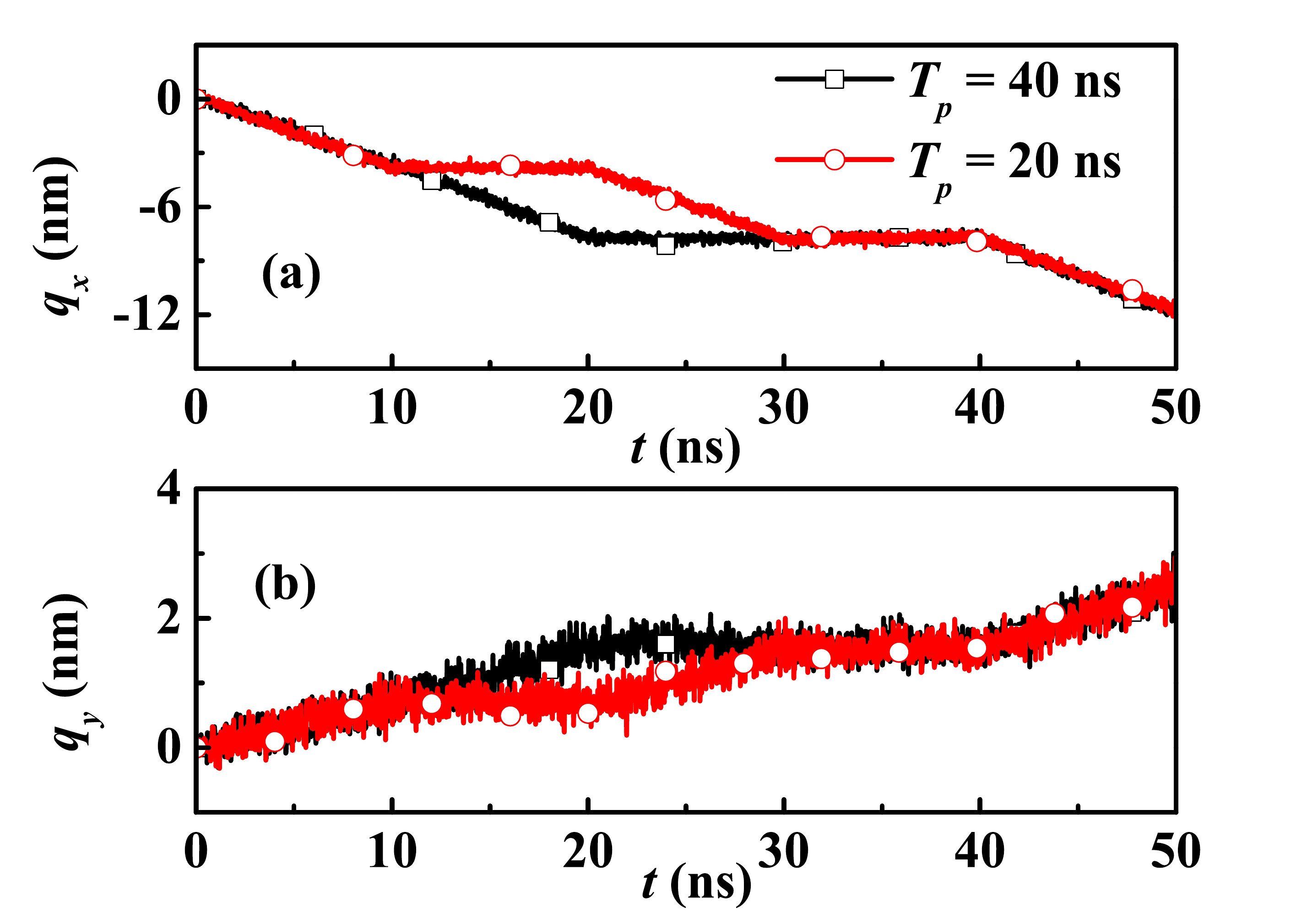}
    \caption{\label{time-pulse}  Under time-varying periodic rectangular pulses (i.e., in every single period $ T_p $, located electric fields with $ E_y = \pm 0.15 \mathrm{MV/cm} $ are applied in a half period $ n T_p < t < n T_p + T_p/2$, ) the motion of the center of skyrmion ($ q_x, q_y $) in time are shown in the $ x $ (a) and $ y $ (b) directions. Temperature is equal to $ T = 25K $.}
\end{figure}

The energy balance of the magnonic spin current depends on two main factors: Due to the phenomenological Gilbert damping, the driven magnonic system continuously loses the energy.
On the other hand the ferroelectric polarization $\vec{P} =  c_{E} [(\vec{m} \cdot \nabla) \vec{m} - \vec{m} (\nabla \cdot \vec{m}) ]$ and the ME coupling with the external electric field  $ E_{ele} = - \vec{E} \cdot \vec{P} $ supplies a ferroelectric energy to the magnonic subsystem (eventually through the energy needed to generate and  maintain $\vec{E}$). Thus, the Gilbert damping plays the role of a sink, and the ferroelectric polarization plays the role of the energy reservoir that sustains the spin current.  To prove this scenario, instead of the constant electric field, we applied a series of rectangular pulses of the electric field, i.e., the electric field is periodically switched on and off. As we see, when the electric field is switched off, the skyrmion stops moving Fig. \ref{time-pulse}. The same behavior we observe for the magnonic spin current (not shown), the current disappears soon after the electric field is switched off. After the electric field is switched off, due to the Gilbert damping, the ferroelectric energy becomes zero. Switching on the electric field pumps the ferroelectric energy into the system and restores the ferroelectric torque. Thus, an electric field has to continuously supply the ferroelectric energy to the magnetic system in both cases, either for restoring the ferroelectric term or for compensating energy losses due to damping. For exploring the microscopic mechanisms of the energy exchange we utilize the quantum Hamiltonian and apply the external electric field $E=\big(0, E_{y}, E_{z}\big)$.  We find

\begin{eqnarray}
        \label{quantumHamiltonian1}
         && \hat{H}=\hat{H}_{ex}+\hat{H}_{D},\nonumber\\
         && \hat{H}_{ex}=J\sum_{n}\hat{S}_{n}\hat{S}_{n+1},\nonumber\\
         && \hat{H}_{D}=-2D_{z}\sum_{n}\hat{z}\cdot\big(\hat{S}_{n}\times\hat{S}_{n+1}\big)-\nonumber\\
         &&  2\sum_{n}D_{y,n}\hat{y}\cdot\big(\hat{S}_{n}\times\hat{S}_{n+1}\big).
    \end{eqnarray}
Here, the effective DMI constants are defined as follows $ D_{z} = E_{z} J e a / 2E_{so} $, $ D_{y}\big(x\big) = E_{y}\big(x\big) J e a / 2E_{so} $ and we assumed that the field applied  along the $\vec{\hat{y}}$ axis is inhomogeneous in the  $\vec{\hat{x}}$ direction. Utilizing  the Holstein-Primakoff transformation we  derive the equation that quantifies the energy exchange between the magnonic $\hat{H}_{mag}=2SJ\sum_{n}\hat{a}^{\dag}_{n}\hat{a}_{n}$
and the ferroelectric subsystems $\hat{H}_{D}=\frac{2S}{i}\sum_{n}D_{z,n}\big(\hat{a}_{n}\hat{a}^{\dag}_{n+1}-\hat{a}_{n}^{\dag}\hat{a}_{n+1}\big)+
2S\sum_{n}D_{y,n}\big(\hat{a}^{\dag}_{n}\hat{a}_{n}\big(\hat{a}^{\dag}_{n+1}+\hat{a}_{n+1}\big)-\hat{a}^{\dag}_{n+1}\hat{a}_{n+1}\big(\hat{a}^{\dag}_{n}+\hat{a}_{n}\big)\big)$, (here $J$ is the exchange constant, $S$ is the spin of the electron and $\hat{a}^{\dag}_{n},~~\hat{a}_{n}$ are the magnon creation and annihilation operators). After standard algebraic transformations we obtain

\begin{eqnarray}
        \label{quantumHamiltonian2}
         && \frac{d}{dt}\langle\hat{H}_{mag}\rangle=i[\hat{H}_{D},\hat{H}_{mag}]=4S^{2}J\times\big\langle \sum_{n}i\big(\hat{a}^{\dag}_{n}-\hat{a}_{n}\big)\cdot \nonumber\\
         && \big(D_{y,n}\hat{a}^{\dag}_{n+1}\hat{a}_{n+1}-D_{y,n-1}\hat{a}^{\dag}_{n-1}\hat{a}_{n-1}\big)\big\rangle .
    \end{eqnarray}

On the right-hand side of the Eq.(\ref{quantumHamiltonian2}) we have the product of two operators. The expectation value of the first operator $\langle i\big(\hat{a}^{\dag}_{n}-\hat{a}_{n}\big)\rangle$
is nonzero at nonzero temperature, while the expectation value of the second operator $\langle\big(D_{y,n}\hat{a}^{\dag}_{n+1}\hat{a}_{n+1}-D_{y,n-1}\hat{a}^{\dag}_{n-1}\hat{a}_{n-1}\big)\big\rangle$ is nonzero
if the translation symmetry in the system is broken, meaning when the  inhomogeneous electric field is applied
$n\mapsto n+2,~~\hat{a}^{\dag}_{n-1}~\hat{a}_{n-1}\mapsto\hat{a}^{\dag}_{n+1}\hat{a}_{n+1}$ but $D_{y,n+1}\neq D_{y,n}$. Thus,  in order to have energy flow from the ferroelectric subsystem to the magnonic subsystem and to
sustain the magnonic spin current one needs simultaneously  two ingredients: applied thermal bias and a nonuniform electric field. We note that in the standard spin Seebeck experiments, translation invariance is
broken by the spatially  non-uniform temperature profile, while in our case temperature profile is uniform and translation symmetry is broken by the nonuniform electric field.

\section{Conclusions}

Our motivation in the present work has been to explore new ways of controlling the motion of skyrmions in a thin ferromagnetic insulator film.
We find that a spatially inhomogeneous electric field serves this purpose in that it leads to a specific torque capable of dragging the skyrmion even at a zero temperature, while thermal effects assist the skyrmion drag and enhance skyrmion velocity. Electric fields are advantages in several ways. They can be generated and temporally controlled in a versatile manner. The spatial inhomogeneities can also be well designed, for instance by nanopatterning the sample by a metallic shielding that blocks the electric field on a microscopic scale.  Generally, the current findings point to an interesting and rich spin-current physics of skyrmionic systems driven by local THz fields.

\section{Acknowledgment}
We acknowledge financial support from DFG through SFB 762 and SFB TRR227, the National Natural Science Foundation of China (No. 11704415, 11674400, 11374373), and the Natural Science Foundation of Hunan Province of China (No. 2018JJ3629, 13JJ2004).

\end{document}